\documentclass[10pt,journal,cspaper,compsoc]{IEEEtran}

\usepackage{amsmath}
\usepackage{url}
\usepackage{enumerate}
\usepackage[pdftex]{graphicx}
\usepackage{psfrag}
\usepackage{setspace}
\usepackage{tabu}
\usepackage{algpseudocode}
\usepackage{algorithm}
\usepackage{caption}
\makeatletter
    \newcommand{\thickhline}{%
        \noalign {\ifnum 0=`}\fi \hrule height 1pt
        \futurelet \reserved@a \@xhline
    }
    \newcolumntype{"}{@{\vrule width 1pt}}
    \makeatother
    \newcolumntype{L}{>{\arraybackslash}m{7cm}}
\ifCLASSOPTIONcompsoc
\else
\fi
\begin{document}
\author{\large Parisa ~Rahimzadeh and~Farid~Ashtiani,~\IEEEmembership{Member,~IEEE}
\\
\IEEEcompsocitemizethanks{\IEEEcompsocthanksitem The authors are with the Department of Electrical Engineering and the Advanced Communications Research Institute (ACRI), Sharif University of Technology, Azadi st., PO: 11155-4363, Tehran, Iran.\protect\\
E-mail: rahimzadeh@ee.sharif.edu, ashtianimt@sharif.edu
}
\thanks{}}
\title { \LARGE Analytical Evaluation of Saturation Throughput of a Cognitive 802.11-based WLAN Overlaid on a WiMAX-TDD Network}


\newcommand*{\factor}{1}
\IEEEcompsoctitleabstractindextext{
\begin{abstract}
This paper analyzes the saturation throughput of a cognitive single hop WLAN overlaid on a primary IEEE 802.16e TDD WiMAX network. After the contention among the secondary nodes, the winner node transmits its data packet in the empty slots of downlink subframes of WiMAX. Regarding the OFDMA structure as well as time-scheduled resources in WiMAX, the time duration of opportunities for the secondary network does not follow simple exponential on-off pattern. To model the dynamic behavior of opportunities for secondary nodes as well as contentions to exploit the opportunities, we propose an analytical model comprised of a discrete-time Markov chain and two inter-related open multi-class queueing networks. The effects of random number of empty slots at different frames as the result of random amount of download data, random packet transmission time at WLAN due to random opportunities in different frames, the dependency of the number of empty slots at consecutive WiMAX frames, and the details of 802.11 MAC protocol are included in our analytical approach. We compare the effect of two resource allocations, i.e., horizontal and vertical striping on the saturation throughput of the cognitive WLAN. Simulation results confirm the accuracy of our analytical approach in different conditions.\\
   
\end{abstract}

 }
\maketitle
\captionsetup{font=small}
\section{Introduction}

 \IEEEPARstart{W}{ith} the increasing popularity of wireless services, efficient use of frequency spectrum becomes more important. Cognitive radio (CR) technology is able to help in efficient use of the spectrum. In this respect, the spectrum assigned to legitimate users (called primary users (PUs)) is used by a second group of users (called secondary or cognitive users (SUs/CUs)) in different methods, e.g., overlay structure \cite{R3}. In a typical overlaid cognitive network, SUs dynamically utilize empty parts of the spectrum (i.e., spectrum holes) \cite{R1}. A part of the spectrum is recognized as a spectrum hole by an SU if it is not assigned to PUs, the corresponding PUs are inactive temporarily, or it is used by PUs very far from the SU. Until now different aspects of CR technology including physical layer and MAC layer issues have been studied in the literature with different scenarios for primary and secondary networks. 

Since the cognitive nodes have to be able to sense and monitor the radio channels occupancy, one of the important issues in physical layer topics is the sensing process in secondary nodes. The sensing process strongly depends on the activity pattern of primary users. One of the consequences of sensing process is spectrum mobility \cite{R3}. That is, an SU that opportunistically uses a frequency channel needs to change that channel when the corresponding channel is sensed as active. Considering several problems in the sensing process, it would be useful to predict the spectrum usage of primary users. In this respect, exploiting the information of time-frequency allocation in the primary network is a very efficient method. Fortunately, in some standards like WiMAX and LTE, occupancy status of resource units is broadcast periodically \cite{R5}, so with secondary nodes able to detect this signal, no sensing is needed. On the other hand, regarding the wide usage of such wireless network standards, they can be suitable choices for primary networks. One of the features of the aforementioned mobile standards (i.e., WiMAX, LTE) reverts to their regular frame structure, such that the time-frequency allocation maps are updated and broadcast to all users at each frame. Another feature of these standards is that they are based on OFDMA. In fact, due to several advantages of OFDMA (e.g., providing high spectral efficiency, excellent coverage, higher data rates, and high performance in fading environments), it has been selected as the resource allocation scheme to users in the new mobile standards \cite{R7}. In this technique, a combination of time and frequency comprises the basic unit of the resources. And based on the user requirements, different number of resource units may be allocated to a specific user at different frames. Hence, although the usual traffic model for the activity status of each frequency channel in primary networks is the on-off traffic model with exponential on and off intervals (e.g., \cite{R8}, \cite{R2}, \cite{R11}), in an OFDMA-based network (e.g., WiMAX or LTE) that resources are allocated to PUs at each frame, the on/off status of PUs are correlated and we need to consider PUs, altogether, frame by frame. Moreover, we should include the memory in resource allocation in consecutive frames. Thus, the previous conventional analyses that usually focus on the status of PUs separately are not applicable anymore.

Up to now some research works have considered OFDMA primary networks in a cognitive network scenario. In \cite{R12} in order to solve the problem of white spots in a WiMAX coverage area (the areas where no network coverage is available for customers), a technique based on the cognitive heterogeneous network is presented. In \cite{R13} an OFDMA primary network like LTE and a pair of secondary transmitter-receiver are assumed. The interference between the primary and secondary nodes is canceled by exploiting the null-space of the channel from the secondary transmitter to the primary receiver at a cost of knowing perfect CSI in the secondary transmitter. In \cite{R14} the OFDMA spectrum occupancy under different traffic models is simulated. Also impacts of different resource unit allocation algorithms in WiMAX (i.e., vertical, horizontal, and rectangular striping) on SUs’ sensing process are discussed. In \cite{R16}, by using cognitive technology, a scheme for operation of an ultra-wideband (UWB) device in WiMAX frequency band is proposed, and detection and avoidance problems in this system are investigated. In \cite{R17} the spectrum sensing issue of the cognitive nodes in the presence of a WiMAX primary network is explored and time and frequency domain spectrum sensing techniques are compared by simulation. In \cite{R18} the power control mechanism and opportunistic interference cancellation in the secondary network is investigated in an underlay spectrum usage model with a WiMAX primary network. In \cite{R19} classification of OFDM signals of the OFDM-based primary networks (e.g., mobile WiMAX and LTE) is investigated. A cognitive radio-based resource allocation in a WiMAX macro-femto two tier network is proposed in \cite{R20}.

In addition to most of the aforementioned works that focus on the sensing process in an OFDMA-based primary networks (e.g., WiMAX), in some papers the transmitted information about the situation of the spectrum holes in the primary OFDMA network is exploited to avoid the sensing process difficulties. In \cite{R26} the authors have proposed two cognitive MAC protocols in the presence of an IEEE 802.16 primary network. In this work, SUs hear the resource allocation maps (DL/UL-MAP) broadcast by the WiMAX base station (BS); hence, they know the empty slots of the coming frame without sensing. Two access protocols (ordinated approach and contention approach) have been evaluated in terms of delay and throughput by simulation. The contention approach is based on CSMA/CA technique where the user with a packet to send waits a random number of frames before the transmission begins. In \cite{R27} an opportunistic spectrum access algorithm for secondary users in the presence of a mobile WiMAX primary network is proposed. In this structure, SUs are aware of spectrum holes by overhearing DL/UL-MAP. Efficiency of the algorithm with a WiMAX primary and secondary network is investigated by simulation. In this scenario, base station of secondary network acts as the secondary decision center, which hears the DL/UL-MAP and according to the proposed algorithm assigns white spaces to SUs. In \cite{R28}, the authors have considered a time-scheduled primary network, e.g., mobile WiMAX, and proposed a simple analytical model to calculate the saturation throughput of a WLAN secondary network overlaid on the primary network. In that scenario, WLAN nodes use empty slots of downlink (DL) subframes of WiMAX to send their packets. Although the analysis in \cite{R28} leads to sufficiently accurate results for FDD mode, in TDD mode the results have non-negligible errors.

From a different point of view, in many papers the coexistence of a cellular network (e.g., WiMAX) and a WLAN in a heterogeneous network has been investigated \cite{R23}, \cite{R24}. However, in our considered scenario, we investigate the coexistence of a WLAN network with a WiMAX network, where the WLAN users opportunistically use the empty slots of DL subframes of WiMAX.

In this paper, we focus on a cognitive network scenario, consisting of an IEEE 802.16e mobile WiMAX primary network in the PMP (point-to-multipoint) mode and a WLAN secondary network that uses the empty slots of DL subframes of WiMAX to send its packets. Secondary nodes use IEEE 802.11 MAC protocol to access the total empty slots (see Fig. 1) in DL subframes. It is also assumed that secondary nodes are able to hear the DL-MAP signal at the beginning of each frame, thus they know the situation of empty slots of DL subframes. Due to the frame structure of the primary network and the method of resource allocation to different primary users, it is not possible to model the activity of individual frequency channels. Instead, we model the dynamic status of all downlink time-frequency resources of the primary network by considering the service requests of all primary users at each frame. To this end, we use a discrete-time Markov chain, which represents the packet arrival process in the buffer at the BS of the primary network as well as the transmission over the limited number of slots in each upcoming frame. Although our modeling approach for primary network is similar to \cite{R28}, there are significant differences in using the components of the Markov chain in our analysis. In fact, due to random nature of packet arrival process in the WiMAX BS, there are random number of empty slots in DL subframes. Moreover, the number of empty slots at consecutive DL subframes is correlated, because if the resources are not sufficient to send all buffered data in WiMAX BS during the upcoming frame, remaining data will be transmitted in next frames. These two subtleties have not been considered in \cite{R28}. Moreover, our analytical approach for the performance evaluation of the secondary network is very detailed compared with the approach in \cite{R28}, leading to more exact results in 802.16e TDD mode (the maximum mismatch error between analytical results and simulations is reduced from more than 18\% to less than 3\%). Also, we have compared the saturation throughput of an SU in our scenario with an algorithm similar to the contention-based MAC algorithm proposed in \cite{R26}. The results show a noticeable improvement in saturation throughput of an SU in our considered scenario.

\begin{figure*}
\centering
\includegraphics[width=6in]{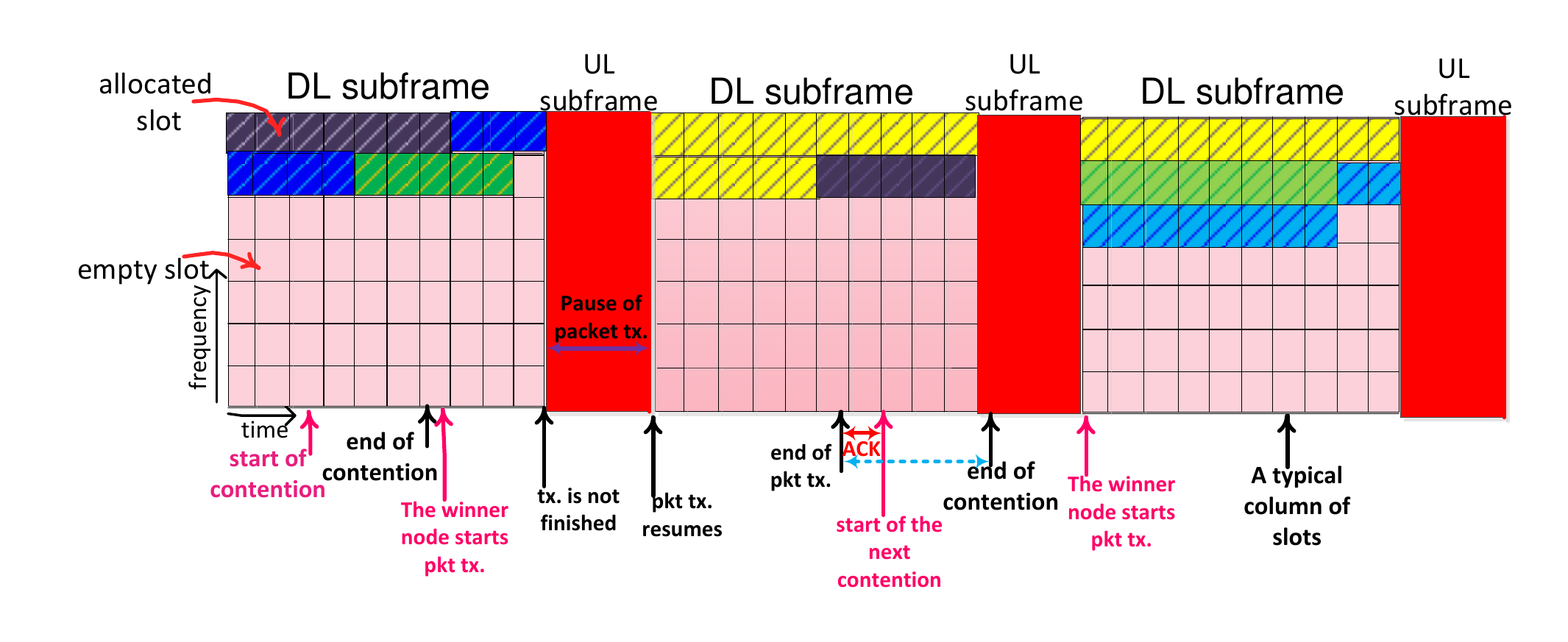}
\caption{Random number of empty slots in the horizontal sriping and an example of time sequence of packet transmission in the secondary network.}
\label{frame}
\end{figure*}

Succinctly, the main contributions of our analysis in this paper can be summarized as in the following:
\begin{enumerate}
  \item Modeling the dynamic nature of resources in DL subframes and including the random number of empty slots in performance evaluation of the secondary network,
  \item Proposing an open multi-class queueing network (called as $QN^{(1)}$) \cite{R6} to pursue the time sequence of transmission process of WLAN packets (including contention among SUs) with high accuracy (see Fig. \ref{frame}),
  \item Proposing another open multi-class queueing network (called as $QN^{(2)}$) to find some time parameters in the previous queueing network, e.g., average packet transmission time for SUs, with respect to random nature of opportunities,  
  \item Deriving the saturation throughput of the cognitive network, i.e., the minimum rate of packet arrivals such that the secondary network is in the border of instability. 
\end{enumerate}

It is worth noting that maximum saturation throughput of a cognitive network scenario is one of the important decision-making parameters in investment on cognitive network development. In fact, it can be used as an important parameter to estimate the maximum number of SUs to be supported with a minimum required throughput.

The rest of the paper is organized as follows. In Section 2 we describe the cognitive network scenario which includes primary and secondary network characteristics. Section 3 presents the proposed analytical model for the primary network. In Section 4 the analytical model for the secondary network is described which consists of the queueing network representing the packet transmission process and the queueing network for derivation of the average packet transmission time. Section 5 is dedicated to numerical results to show the accuracy of the proposed model compared to simulations and Section 6 concludes the paper.
\section{cognitive network scenario}
In this section, we explain the characteristics of primary and secondary networks. We consider a cognitive network consisting of an IEEE 802.16e WiMAX primary network and cognitive nodes which opportunistically utilize empty slots of DL subframes of WiMAX. In fact, secondary network consists of WLAN nodes that send their packets to the access point (AP) based on IEEE 802.11 MAC protocol using the frequency spectrum holes (empty slots) of WiMAX. 
\subsection{Primary Network}
As stated before, based on several advantages of OFDMA it is used in IEEE 802.16e standard. WiMAX frames are divided into symbols in time domain and orthogonal subcarriers in frequency domain. Slot is the resource allocation unit in IEEE 802.16e, which consists of a subchannel (comprised of a few subcarriers) over a few OFDM symbols \cite{R29}. The arrangement of subcarriers in a subchannel and the number of time symbols in a slot is dependent on the type of OFDMA permutation scheme \cite{R7}. It is assumed that PUSC (partial usage of subchannels) subchannelization method is used in the mobile WiMAX primary network \cite{R14}, \cite{R30}. In PUSC, which is the most commonly used subchannelization method \cite{R31}, subchannels are formed by subcarriers distributed pseudo-randomly over the entire bandwidth. Hence this method of subchannelization can exploit frequency and interference diversity \cite{R5}. Since the subcarriers at each slot are distributed through the whole bandwidth, the average channel quality at different slots is assumed to be the same, leading to the same average number of bits transmittable at each slot. In other words, although the fading state of each subcarrier may be different, regarding the frequency distance between them, their states are considered to be independent of each other, so we can assume a similar average channel condition for each slot of WiMAX. 

Although both TDD and FDD duplexing modes are supported in the standard, due to inefficiency of FDD in asymmetric data services, TDD mode is the preferred one in new mobile WiMAX profiles \cite{R7}. So TDD mode is considered in our scenario. In 802.16e standard, TDD mode is capable of adaptive adjustment of downlink to uplink ratio to support asymmetric downlink/uplink traffic. Usually in wireless networks, the offered traffic load in downlink direction is much more than uplink, hence the size of downlink subframes is usually larger than uplink subframes and consequently there may be more opportunities in DL subframes for secondary nodes. On the other hand, regarding the importance of downlink traffic, we have assumed that uplink subframes are considered to be small as much as possible to be matched with upload requirements of users. Therefore the number of empty slots at UL subframes is considered not to be noticeable. Consequently, UL subframes lead to interruptions in secondary packet transmissions. Regarding a stationary traffic load for uplink and downlink, a fixed DL to UL ratio is considered in our analysis; however, the DL to UL ratio can be flexible and dynamic in WiMAX standard. It is worth noting that in the case of exploiting UL opportunities as well, with respect to different characteristics of DL and UL traffic, we need more details in our modeling approach. It will be explained briefly in Section 4. 

In our scenario, we consider the random nature of opportunities based on horizontal and vertical striping algorithms \cite{R14} as the resource allocation algorithms of WiMAX. Furthermore, in the case of other mapping algorithms, e.g., rectangular one, we assume that the empty slots of a DL subframe are distributed uniformly over the entire DL subframe in time and frequency spectrum.

\subsection{Secondary Network}
In the considered scenario, similar WLAN nodes form the secondary network. The nodes are within the transmission range of each other and the packet generation process at all cognitive nodes are independent and identical. Secondary users contend with each other in a control channel (out of WiMAX frequency band) based on IEEE 802.11 MAC protocol  in a 4-way handshaking mode (RTS/CTS/DATA/ACK) to access the opportunities in DL subframes of WiMAX. In our considered scenario, the backoff and contention processes among the secondary nodes are independent of WiMAX frame structure. This scenario leads to increased spectrum efficiency; while in \cite{R26} a complete frame of WiMAX (with possible free resources) is considered as a backoff unit. Moreover, in our scenario there is no need for synchronization between control signals of SUs (i.e., RTS, CTS, ACK) and WiMAX time symbols.

We focus on the MAC layer issues in this paper, so the necessary physical layer compatibilities as well as physical layer impairments are ignored. Indeed, noting that the winner SU uses empty slots of the DL subframes of WiMAX to send its packets, we have assumed that SUs exploit OFDM technique for their transmissions. So it is possible to use several non-contiguous slots in a transmission. Following each contention, the winner node exploits all empty slots to send its fixed size packets. Since the frame control section of the WiMAX frame which is used to send control information for users is not encrypted \cite{R33}, \cite{R10}, all SUs are able to overhear the DL-MAP signal broadcast at the onset of each frame, so they are aware of empty slots at each DL subframe without any need for a sensing process. In this case PUs and WiMAX base station operation are not affected by SUs activities. Obviously, if DL/UL-MAP is not detectable at SUs (e.g., is encrypted), the scenario needs sensing the opportunities of each frame that is not within the scope of this paper.\\
\vspace{-0.1cm} 
Since we focus only on DL opportunities, if the contention between secondary users ends during a UL subframe, the winner user should wait until the beginning of the next DL subframe to start data transmission in empty slots of that frame. In addition, if sending the packet is not finished in the current frame, it resumes in the next frame (without any contention) after an interruption during the UL subframe. Obviously, dependent upon the number of empty slots of frames and size of the packets, such interruptions may be experienced several times during a packet transmission which should be considered in the packet transmission time of secondary nodes. Consequently, packet transmission time also depends on the time instant when it has begun (see Fig. \ref{frame}). Besides, since the number of empty slots at each DL subframe is dependent upon the amount of PU’s data to be downloaded on one hand, and the starting instant of SU packet transmission is distributed through the DL subframe on the other hand, each packet transmission prolongs a random number of OFDM symbols. Therefore, the bottleneck for the throughput of the secondary network is collision among secondary nodes as well as the limited opportunities in the DL subframe of WiMAX. Thus, analytical modeling of packet transmission is of crucial importance.
In order to facilitate following the analytical model in next sections, a brief description of variables is presented in Table \ref{glossary}.\\
\vspace{-0.7cm}
\begin{table*}[!ht]
  \small
  \centering
  \caption{Description of Symbols}
    \begin{tabular}{{|p{3.2cm}|p{13.1cm}|}}
    \thickhline
    \textbf{\quad Symbol} & \textbf{Description} \\
    \thickhline
    \quad  $P$ & Transition probability matrix of the Markov chain in Fig. \ref{MC}   \\\hline
    \quad $P_{ij}$ & Transition probability from state $i$ to state $j$ of the Markov chain in Fig. \ref{MC} \\\hline
    \quad $P_k$ & The probability of arrival of $k$ packets during a frame to the WiMAX BS \\\hline
    \quad $\lambda_p$ & Average of packet arrival rate to WiMAX BS \\\hline
    \quad $M$ & The total number of slots in DL subframe\\\hline
    \quad $N$ &The number of required slots to transmit all packets of a full WiMAX BS buffer \\\hline
    \quad $S_P$ & The number of required slots to transmit a WiMAX packet \\\hline
    \quad $S_S$ & The number of required slots to transmit a WLAN packet \\\hline
    \quad $C_B$ & Size of the finite buffer at WiMAX BS in number of packets  \\\hline
    \quad $\pi$ & Steady state probability vector of the Markov chain in Fig. \ref{MC}  \\\hline
    \quad $W_0$ & Initial contention window size in the WLAN \\\hline
		\quad $m$ & The number of maximum backoff stages in the WLAN \\\hline
		\quad $P^{(I)},P^{(C)},P^{(S)}$ & Probabilities of idle, collision, and successful transmission time slots, respectively. \\\hline
			\quad $T^{(I)},T^{(C)}$ & Time duration of idle and collision transmission time slots, respectively. \\\hline
		\quad $K_{sym}$ & The number of time symbols over a WiMAX frame\\\hline
		\quad $K_{sym,DL}$ & The number of time symbols over a DL subframe of WiMAX \\\hline
		\quad $r^{u,u'}_{l,l'}$ & The probability that a class-$l$ customer from node $u$ is routed as a class-$l'$ customer to node $u'$ in the open queueing networks in Figs. \ref{Q1} and \ref{Q2} \\\hline
				\quad $r^{u,out}_{l,l'}$ & The probability that a class-$l$ customer from node $u$ is routed as a class-$l'$ customer to out of the queueing networks in Figs. \ref{Q1} and \ref{Q2}\\\hline
				\quad $r^{out,u}_{l}$ & The probability that a class-$l$ customer  is routed to node $u$ from out of the queueing network in Fig. \ref{Q2}\\\hline
				\quad $P_{col}$ & The collision probability for a WLAN user\\\hline
				\quad $\tau^u_l$ & The average service time of a class-$l$ customer at node $u$ in Figs. \ref{Q1} and \ref{Q2}\\\hline
				\quad $T_{RTS},T_{CTS},T_{ACK}$ & The required times to send RTS, CTS, and ACK signal in the control channel, respectively.\\\hline			
				\vspace{0.005cm}
				\quad $\hat{T}$ & \vspace{0.005cm} $T$ in number of WiMAX time symbol\\\hline
				\quad $\nu$ & The number of time symbols in a slot of WiMAX \\\hline
				\quad $f^u(i)$ & a function that denotes the number of empty slots of DL subframe of WiMAX along the frame\\\hline
				\quad $\gamma_i$ & The average time that sending a WLAN packet in the empty slots of WiMAX prolongs, when it has been started at the $i$-th symbol over the WiMAX frame\\\hline
			\quad $\beta_{i,i'}$ & The probability that an SU packet transmission which has been started at the $i$-th symbol over the WiMAX frame ends at the $i'$-th symbol of a frame\\\hline
		\quad $\alpha^u_l$ & The arrival rate of class-$l$ customers at node $u$ in the queueing networks in Figs. \ref{Q1} and \ref{Q2}\\\hline
		\quad $\alpha^{u,out}_l$ & The departure rate of class-$l$ customer from node $u$ to out of the queueing networks in Figs. \ref{Q1} and \ref{Q2}\\\hline
		\quad $\rho$ & The average number of customers at the queueing network in Fig. \ref{Q1}\\\hline
		\quad $\rho^u$ &The average number of customers at node $u$ at the queueing network in Fig. \ref{Q1}\\\hline
		\quad $\lambda_l$ & Class-$l$ customer arrival rate at the queueing networks in Fig. \ref{Q1} and \ref{Q2}\\\hline
		\quad $\lambda$ & Total customer arrival rate at the queueing network in Fig. \ref{Q1}\\\hline
		\quad $e^u$ & Total empty slots in a DL subframe which node $k$ in queueing network in Fig. \ref{Q2} represents\\\hline
		\quad $e_{res}^{u}$ & Residual empty slots of a DL subframe (represented by node $u$) for a packet transmission starting at the $x$-th symbol\\\hline
				\quad $T_{frame}$ & Time duration of a WiMAX frame\\\hline
				\quad $T_{sym}$ & Time duration of a time symbol in a WiMAX frame\\\hline
		\quad $T_{DL}$ & Time duration of the DL subframe in a WiMAX frame\\\hline
			\quad $R$ & Downlink to uplink subframe ratio\\\hline
    \end{tabular}%
  \label{glossary}%
\end{table*}%

\section{analytical model for primary network}
\begin{figure}
\centering
\includegraphics[width=3.3in]{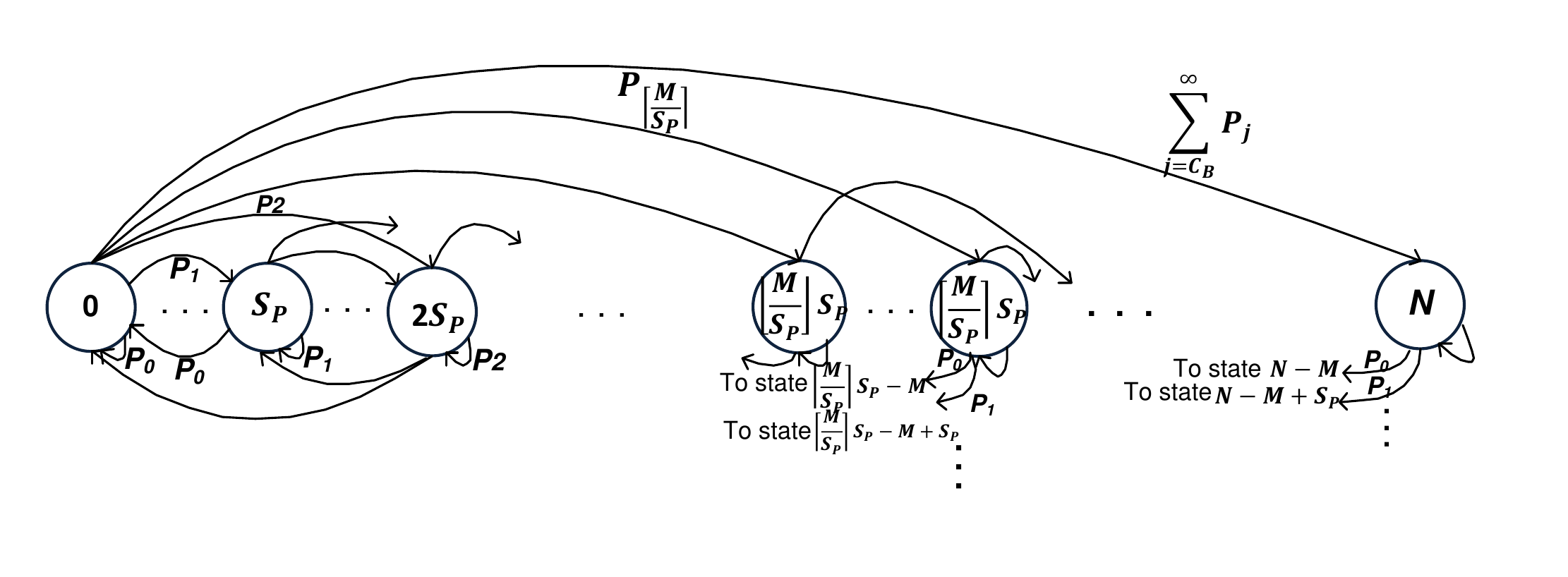}
\caption{Markov chain used for modeling the primary network.}
\label{MC}
\end{figure}

To evaluate the performance of the cognitive network scenario, we need to model the random nature of available resources, i.e., empty slots, in the primary network. As indicated in the previous section, the secondary packets are transmitted through empty slots in DL subframes. So the status of empty slots affects the packet transmission time. In this section, we explain the analytical model employed for modeling the primary network in order to find the statistical properties of the number of empty slots in DL subframes of WiMAX. \\
\vspace{-0.05cm}
Considering the fact that in PUSC subchannelization method, subcarriers of one subchannel is distributed over the entire bandwidth, we have assumed the number of bits transmitted in one slot (called as data unit in this paper) is fixed. It is also assumed that the fixed size packets (comprised of $S_p$ data units) of all PUs arrive at a limited buffer located at BS, with a Poisson distribution. In this case by arrival of one packet at the BS there is a batch increase in the number of required slots in the upcoming frame. Therefore, we are able to consider a discrete time Markov chain (DTMC) (Fig. \ref{MC}) in order to represent the status of empty slots at DL subframes. In \cite{R28}, arrival of data units is assumed to be Poisson which is a simplified and non-practical assumption. It is worth noting that considering more general arrival processes such as batch Markov arrival process for packets can be included in our model; however, for the sake of simplicity and focusing on the interactions between secondary and primary networks a Poisson process is considered for the packet arrivals at the primary network.

In the Markov chain in Fig. \ref{MC}, each state represents the number of slots required to transmit the packets stored in the BS buffer at the beginning of a frame. The number of packets in the buffer at the beginning of a specific frame is dependent on the number of recently arrived packets during the last frame as well as the number of packets existed in the buffer at the beginning of the last frame. By solving global balance equations (GBE) of the Markov chain, we are able to derive the probability mass function (pmf) of the number of empty slots in DL subframes. GBE of this Markov chain is written as
\begin{align}\label{gbe1}
\pi P=\pi ,
\end{align}
for $i=0,..., \lfloor\frac{M}{S_P}\rfloor S_P :$
\begin{align}\label{gbe2}
\begin{split}
\left\{ 
\begin{array}{l l}
P_{i,kS_P}=P_k;&\scriptstyle k=0,...,C_B-1\\
P_{i,N}=\sum_{j=C_B}^{\infty}P_j 
\end{array}
\right.,
\end{split}
\end{align}
for $i=\lceil\frac{M}{S_P}\rceil S_P,..., N:$
\begin{align}\label{gbe3}
\begin{split}
\left\{ 
\begin{array}{l l}
P_{i,i-M+kS_P}=P_k;&\scriptstyle k=0,...,\lfloor\frac{N+M-i}{S_P}\rfloor; \\
&\scriptstyle i-M+kS_P\neq N,\\
P_{i,N}=\sum_{j=\lceil\frac{N+M-i}{S_P}\rceil}^{\infty}P_j ,
\end{array}
\right.
\end{split}
\end{align}
where $P$ is the transition probability matrix of the Markov chain, $P_{i,j}$ is the transition probability from state $i$ to state $j$, $P_k$ is the probability of arrival of $k$ packets during a frame, $S_P$ is the number of slots needed for the transmission of a primary packet and $\pi$ represents the steady state probability vector. $N$ equals $C_BS_P$ where $C_B$ is the size of the finite buffer at BS in number of packets and finally $M$ is the total number of slots in a DL subframe. Obviously, if the number of packets arriving at a frame is very large, packet overflow occurs and some of them will be blocked. After solving (\ref{gbe1}), the steady state probability mass function for the buffer status is derived. Since the number of empty slots at consecutive frames are dependent, exploiting the steady state probability of the above Markov chain cannot be employed efficiently. In the next section, we include the structure of the above Markov chain (consisting of the state transition probabilities) in a multi-class open queueing network, to derive the secondary packet transmission time.

It is worth noting that the arrival of a fixed size packet to the buffer of the WiMAX BS causes a batch increase in the allocated slots of the coming frames. Thus, we could include variable size packets for WiMAX by changing the transition probabilities of the proposed Markov chain according to the probability mass function for the number of slots the batch requires to be transmitted.

\section{Analytical Model for Secondary Network}
\begin{figure}
\centering
\includegraphics[width=3.3in]{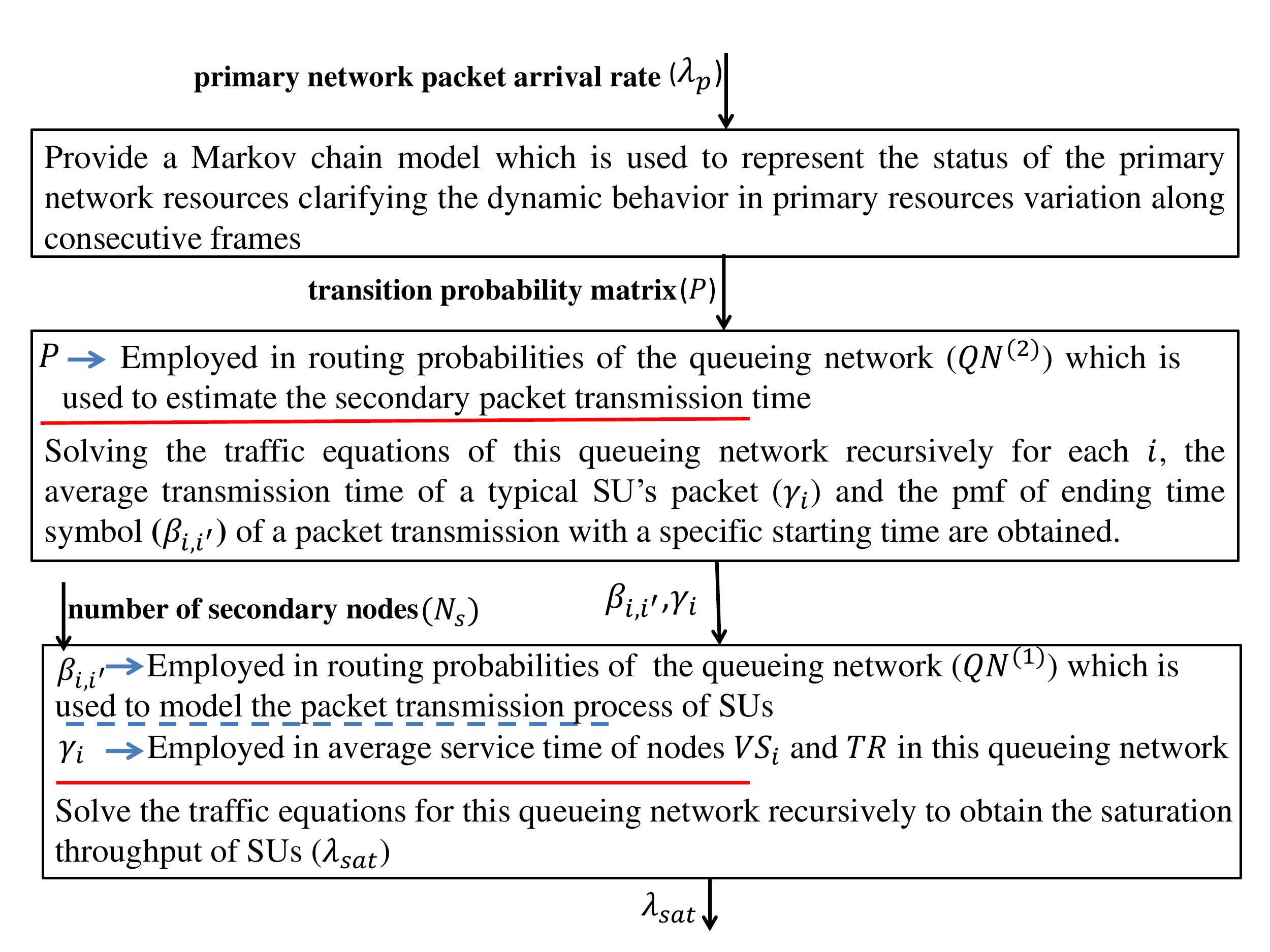}
\caption{The sequence of analysis stages for deriving the saturation throughput of secondary network.}
\label{flowchart}
\end{figure}

In this section, we present our analytical approach to model the behavior of the secondary WLAN nodes and packet transmission process at SUs. First, we explain our proposed multi-class open queueing network ($QN^{(1)}$) used to model the packet transmission process in WLAN nodes based on IEEE 802.11 MAC protocol. Second, in order to find some parameters of this network (some routing probabilities and service times), we propose another queueing network ($QN^{(2)}$) that its parameters include the transitions among the states of the Markov chain corresponding to primary network. A flowchart explaining the overall proposed analytical model to obtain the saturation throughput of the secondary network is presented in Fig. \ref{flowchart}.

\subsection{Proposed Queueing Network Representing the Packet Transmission Process}
\begin{figure}
\centering
\includegraphics[width=3in]{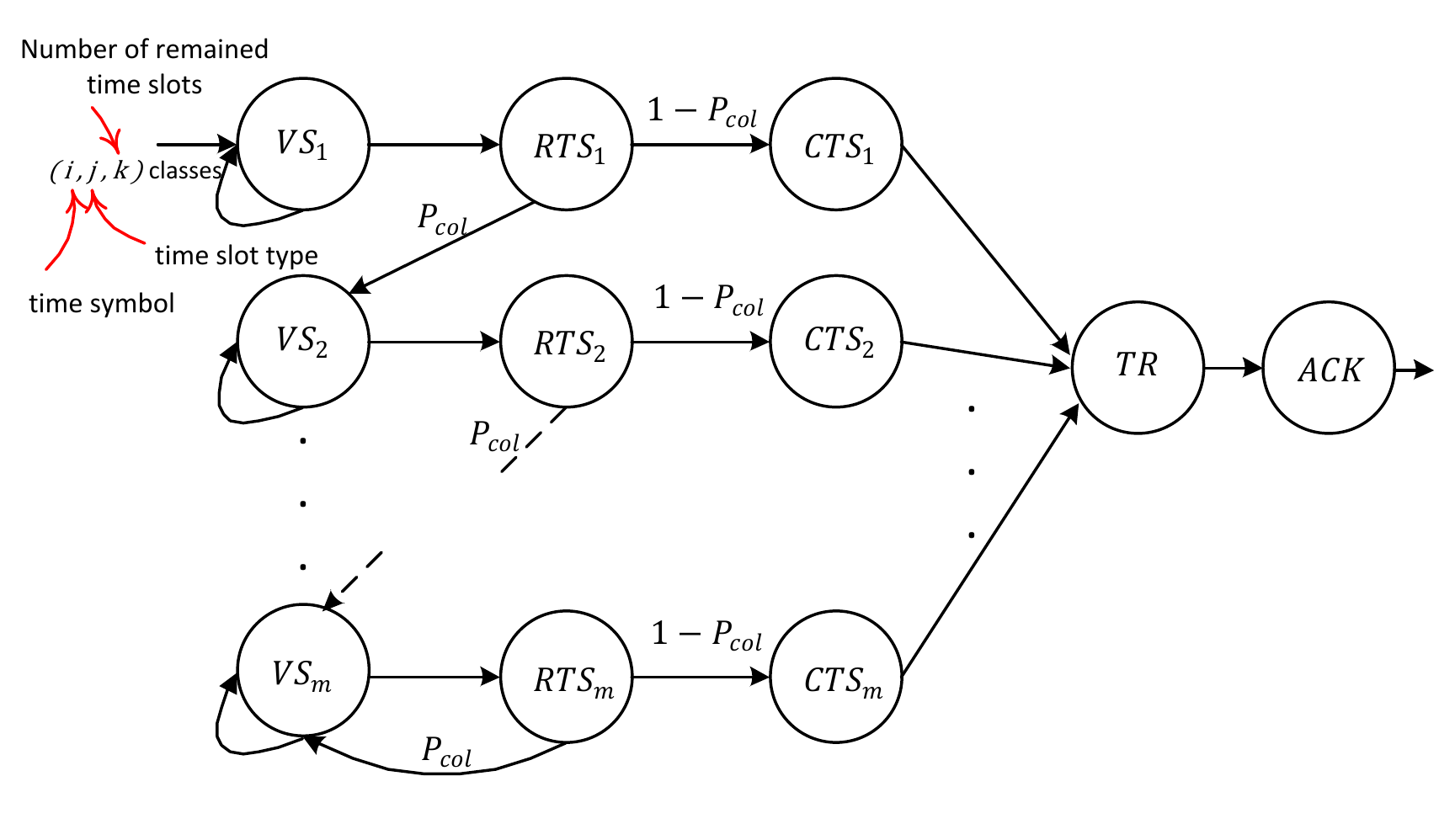}
\caption{Open queueing network model ($QN^{(1)}$) representing the packet transmission process at a typical SU.}
\label{Q1}
\end{figure}

We propose an open multi-class queueing network to represent the behavior of a typical secondary user as in Fig. \ref{Q1}.  This queueing network is in fact an extended version of the model used in \cite{R2}, \cite{R28}. Each node of this queueing network characterizes a stage in the packet transmission process with respect to details of IEEE 802.11 MAC protocol (i.e., backoff, RTS/CTS, data transmission, and ACK) \cite{R32}. In the assumed MAC protocol a user who has a packet to send, sets a random number uniformly distributed in $[0,W_0-1]$ and after each time slot\footnote{To avoid confusion between ‘slot’ in WiMAX network and ‘slot’ in IEEE 802.11 MAC protocol, we use the term ‘time slot’ when 802.11 MAC protocol is considered (in fact, as indicated in Section 2.1, slot in WiMAX is the combination of a number of subcarriers and a number of OFDM time symbols).} it counts down if RTS signal is not sent by another SU. Otherwise the downcounting resumes when the last transmission process has completed, i.e., CTS time-out happens or ACK signal is sent. When the counter reaches zero, the backoff stage ends then RTS signal is sent. If RTS signal is collided with another RTS signal, the contention window size is doubled. And after $m$ collisions the contention window size remains constant. If RTS signal is sent successfully, CTS signal is sent back and sending the packet in empty slots of WiMAX frame begins.

In the proposed queueing network each node is an M/G/$\infty$ node \cite{R21} (see Fig. \ref{Q1}). Customers in this queueing network are SU packets that have to be transmitted. Node $VS_n$ models the time spent in each virtual time slot corresponding to $n$-th backoff stage. In 802.11 MAC protocol three types of time slots, i.e., idle ($I$), collision ($C$), and successful transmission ($S$), occur with probability $P^{(I)}$,$P^{(C)}$, and $P^{(S)}$, respectively \cite{R32}. Since the secondary nodes are considered to be saturated, parameters $P^{(I)}$,$P^{(C)}$, and $P^{(S)}$  are computed similar to \cite{R28} and \cite{R32}.

Customers entering node $VS_n$ have triple classes $(i,j,k)$. $i$ represents time symbol, i.e., it shows number of current time symbol over the frame. Thus, $i$ can be between 1 and $K_{sym}$, where $K_{sym}$ is the number of time symbols over a WiMAX frame. In fact $i$ enables us to trace the regular time structure of the frames. $j$ indicates the type of virtual time slot (i.e., $I,C,S$). Finally, $k$ is a counter that indicates the number of remaining time slots for the backoff counter to reach zero. Routing probabilities and average service times of different classes of customers in node $VS_n$ are written as ($1\leq n\leq m,j=I,C,S,1\leq i,i'\leq K_{sym},1\leq k\leq W_{n-1}-1$)
\begin{equation}\label{r1}
r^{VS_n,VS_n}_{(i,I,k),(i+\hat{T}^{(I)},j,k-1)}=P^{(j)} ,
\end{equation}
\begin{equation}\label{r2}
r^{VS_n,VS_n}_{(i,C,k),(i+\hat{T}^{(C)},j,k-1)}=P^{(j)} ,
\end{equation}
\begin{equation}\label{r3}
r^{VS_n,VS_n}_{(i,S,k),(i',j,k-1)}=P^{(j)}\beta_{i+\hat{T}_{RTS}+\hat{T}_{CTS},i'-\hat{T}_{ACK}} ,
\end{equation}
\begin{equation}\label{T1}
\tau^{VS_n}_{(i,I,k)}=T^{(I)} ,
\end{equation}
\begin{equation}\label{T2}
\tau^{VS_n}_{(i,C,k)}=T^{(C)} ,
\end{equation}
\begin{equation}\label{T3}
\tau^{VS_n}_{(i,S,k)}=T_{RTS}+T_{CTS}+\gamma_{i+\hat{T}_{RTS}+\hat{T}_{CTS}}+T_{ACK},
\end{equation}
where $W_n=2^n W_0$, $r_{l,l'}^{u,u'}$ is the probability that a class-$l$ customer from node $u$ is routed as a class-$l'$ customer to node $u'$, and $\tau_l^u$ is the average service time of a class-$l$ customer in node $u$. $T^{(I)}$ and $T^{(C)}$ are time duration of idle and collision time slots, respectively \cite{R28}. Moreover, $T_{RTS}$, $T_{CTS}$, and $T_{ACK}$ are the times needed to send RTS, CTS, and ACK signals in the control channel, respectively. $\hat{T}$ is $T$ in number of WiMAX time symbols. $\gamma_i$ is the average time that sending a secondary packet in the empty slots of the primary network prolongs, when the transmission has been started at the $i$-th symbol over a WiMAX frame. And $\beta_{i,i'}$ is the probability that an SU packet transmission which has been started at the $i$-th symbol over a WiMAX frame ends at the $i'$-th symbol of the same or another frame. Since secondary wireless nodes are assumed to be saturated, following the end of packet transmission process at each node, the transmission process (including backoff stages, RTS/CTS, data, and ACK) of another packet begins instantaneously. Moreover, routing probabilities and service times not included in (4)-(9), equal zero.\\
\vspace{-0.07cm}
Service time of node $RTS_i$ is the time needed to send RTS packet in the control channel. Upon successful transmission of RTS packet, CTS packet is sent back via receiver and transmission of packets in the remaining empty slots of the current frame begins provided that the current time instant is not within a UL subframe; otherwise, packet transmission starts at the onset of the next DL subframe (see Fig. \ref{frame}). Finally sending ACK signal by receiver indicates the end of packet transmission. \\
\vspace{-0.07cm}
The customers entering nodes $RTS_i$, $CTS_i$, $TR$, and $ACK$ are single-class customers such that their classes represent corresponding time symbol of their arrival instants. The routing probabilities in these nodes can be written as
\begin{align}
\label{r4}
r^{VS_n,RTS_n}_{(i,j,k),(i)}=\left\{
\begin{array}{l l}
1\quad;\text{$k=0$}\\
0\quad;\text{o.w.}
\end{array}
\right.&;
\quad\scriptstyle 1\leq n\leq m,
\end{align}
\begin{align}
\label{r5}
r^{RTS_n,CTS_n}_{(i),(i+\hat{T}_{RTS})}=1-P_{col}&;\quad \scriptstyle 1\leq n\leq m,
\end{align}
\begin{align}
\label{r6}
r^{RTS_n,VS_{n+1}}_{(i),(i+\hat{T}_{RTS},j,k)}=P_{col}P^{(j)}\frac{1}{W_n}&;
\begin{array}{l l}
\scriptstyle
1\leq n\leq m-1\\
\scriptstyle
0\leq k\leq W_n-1
\end{array},\\
\label{r7}
r^{CTS_n,TR}_{(i),(i+\hat{T}_{CTS})}=1&; \quad\scriptstyle 1\leq n\leq m,\\
\label{r8}
r^{RTS_m,VS_{m}}_{(i),(i+\hat{T}_{RTS},j,k)}=P_{col}P^{(j)}\frac{1}{W_{m-1}}&;\quad\scriptstyle 0\leq k\leq W_{m-1}-1,\\
\label{r9}
r^{TR,ACK}_{(i),(i')}=\beta_{i,i'}&,\\ 
\label{r10}
r^{ACK,out}_{(i),(i+\hat{T}_{ACK})}=1&,
\end{align}
where in (\ref{r4})-(\ref{r10}) $j=I,C,S$, $1\leq i,i'\leq K_{sym}$, $r_{l,l'}^{u,out}$ is the probability that a class-$l$ customer from a typical node $u$ is routed as a class-$l'$ customer to out of the queueing network, i.e., when the packet transmission process completes, the corresponding time symbol is $l'$, and $P_{col}$ is the probability of collision among SUs, which is obtained similar to \cite{R28} and \cite{R32}. 

In addition to service times of nodes $VS_n$, the average service times of different classes in the other queueing nodes can also be written as ($1\leq n\leq m, 1\leq i\leq K_{sym}$)
\begin{align}\label{T15}
& \tau^{RTS_n}_{i}=T_{RTS},\quad\tau^{CTS_n}_{i}=T_{CTS},\\
& \nonumber\tau^{TR}_{i}=\gamma_i,\quad\quad\quad\tau^{ACK}_{i}=T_{ACK}.
\end{align}
Due to deterministic service time of customers in nodes $RTS_i$, $CTS_i$, and $ACK$, the corresponding average service times are obtained straightforwardly. However in node $TR$, each packet transmission with different starting time instants will have different average time duration due to UL interrupts and random number of resources. The random nature in number of empty slots in WiMAX frames and its correlation in consecutive frames would cause some complexities. Thus, to calculate $\gamma_i$ and $\beta_{i,i'}$ for all $i$, $i'$ we propose another queueing network that is explained in detail in the next subsection.

By solving traffic equations of $QN^{(1)}$, we are able to find the arrival rate of customers with different classes at each node. Thus, regarding the average service time at each node for different customers, we will be able to derive the traffic intensity (i.e., product of arrival rate by the average service time \cite{R21}) of each node equivalent to the average number of customers at that node. Considering the saturation status of secondary nodes, traffic equation for node $VS_1$ is written as ($1\leq i\leq N_{sym}, j=I,C,S, 0\leq k\leq W_0-1$)
\begin{equation}\label{TEQ1}
\alpha_{(i,j,k)}^{VS_1}=\lambda_{(i,j,k)}+\sum_{(i',j',k')}\alpha_{(i',j',k')}^{VS_1}r_{(i',j',k'),(i,j,k)}^{VS_1,VS_1} ;
\end{equation}
\begin{align}\label{TQ}
\lambda_{(i,j,k)}=P^{(j)}&(\frac{1}{W_0}\alpha_i^{ACK,out}),
\end{align}
where $\alpha_l^u$ denotes the arrival rate of class-$l$ customers at node $u$. The first term in (\ref{TEQ1}) shows the rate of arrivals from exogenous world with time class $(i,j,k)$, which due to saturation is equal to the rate of outgoing customers from the queueing network at the time instant $i$ as in (\ref{TQ}), where $\alpha_l^{u,out}$ denotes the departure rate of class-$l$ customer from node $u$ to out of the queueing network, i.e.,
\begin{align}
\alpha_l^{u,out}=\sum_{l'}\alpha_{l'}^{u}r_{l',l}^{u,out} .
\end{align}
 And for the other nodes in general we have
\begin{equation}
\alpha_l^u=\sum_{u'}\sum_{l'}\alpha_{l'}^{u'}r_{l',l}^{u',u} .
\end{equation}
Note that the traffic equations of this queueing network have to be solved iteratively, considering the dependency between the arrival rate of customers from exogenous world to the queueing network ($\lambda_{(i,j,k)}$) and the rate of outgoing customers from the queueing network.\\
\vspace{-0.05cm}
Using $M/G/\infty$ nodes indicates that different packets at each SU may be sent in parallel. Since in reality, the packets are transmitted one-by-one at each SU and $QN^{(1)}$ represents the packet transmission process, we confine the packet arrival rate at the queueing network $(\lambda=\sum_{(i,j,k)}\lambda_{(i,j,k)})$ such that the average number of customers at the queueing network be kept smaller than one. Thus,
\begin{equation}
\rho=\sum_{n=1}^{m}(\rho^{VS_n}+\rho^{RTS_n}+\rho^{CTS_n})+\rho^{TR}+\rho^{ACK}<1 ,
\end{equation}
where $\rho^{VS_{n}}$, $\rho^{RTS_{n}}$, $\rho^{CTS_{n}}$, $\rho^{TR}$, and $\rho^{ACK}$ denote the traffic intensities of nodes $VS_n$, $RTS_n$, $CTS_n$, $TR$, and $ACK$, respectively. The traffic intensity of a typical node $u$ in the queueing network can be expressed as
\begin{equation}
\rho^u=\sum_{l'}\alpha_{l'}^u\tau_{l'}^u.
\end{equation}
\vspace{-0.05cm}
Since all $N_s$ cognitive nodes are similar and the queueing network in Fig. \ref{Q1} represents a typical cognitive node, in order to find the saturation throughput of the secondary network, the average number of customers in the queueing network is considered equal to one. Thus, the saturation throughput is the minimum rate of packet arrivals at a typical cognitive node 
such that the queueing network is saturated, that is,
\begin{equation}\label{main}
\lambda_{sat}=min\lambda\mid_{\rho=1}.
\end{equation}
Obviously, the saturation throughput of the secondary network equals $\Lambda_{sat}=N_s\lambda_{sat}$.\\
\vspace{-0.05cm}
In addition, by solving the traffic equations we are able to analytically estimate the probability mass function of the starting time instant of packet transmission at a typical frame which is proportional to the arrival rate of customers of different classes at node $TR$. It is worth noting that in \cite{R28} the distribution of the starting time of packet transmission over a frame is assumed to be uniform, which is not necessarily true. \\
\vspace{-0.6cm}
\subsection{Proposed Queueing Network for Derivation of Packet Transmission Time}
As explained before, in order to evaluate the saturation throughput of SUs, we need to find the service time of different nodes and routing probabilities in $QN^{(1)}$. In this part, we compute the parameters $\gamma_i$ and $\beta_{i,i'}$ in (\ref{r3}), (\ref{r9}), and (\ref{T15}). $\gamma_i$ denotes the average transmission time of SU packets, when the transmission has started at the $i$-th time symbol over a frame and $\beta_{i,i'}$ is the probability that the packet transmission started at the $i$-th time symbol, ends at the $i'$-th time symbol (not necessarily in the same frame). 

\begin{figure*}
\centering
\includegraphics[width=6in]{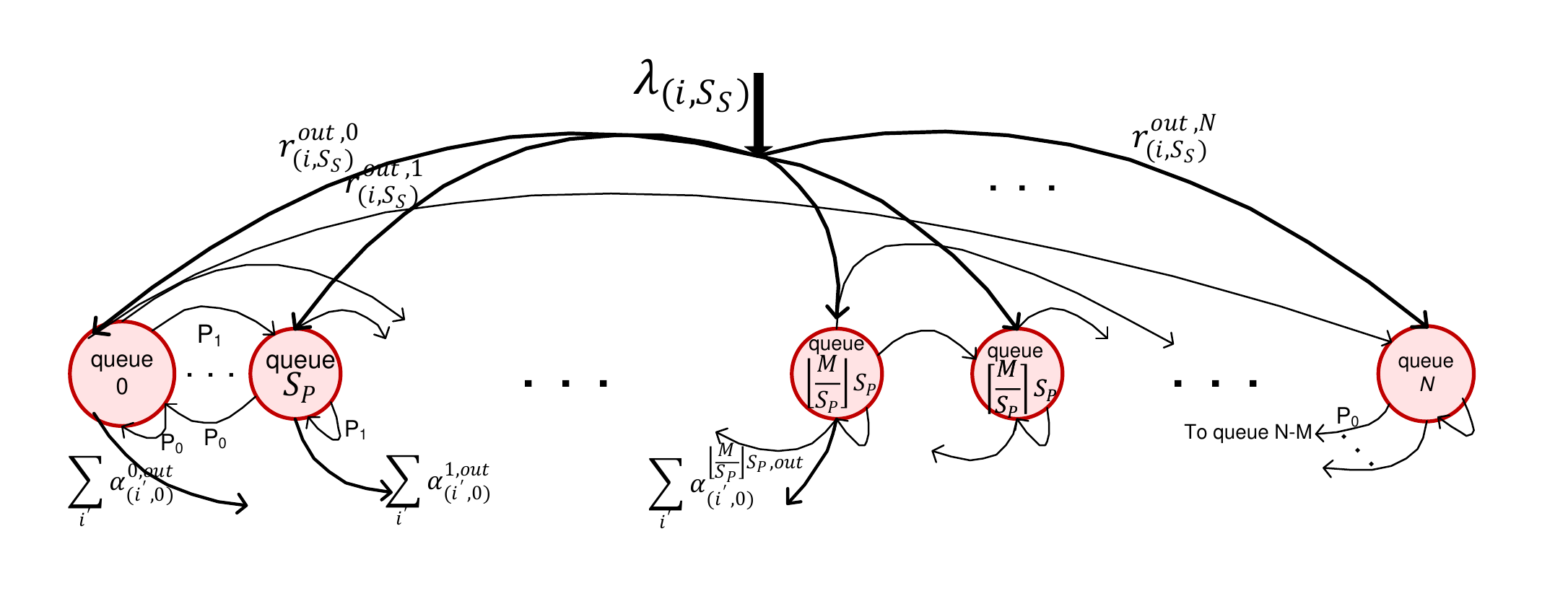}
\caption{The proposed multi-class queueing network for transmissions starting at the $i$˗th symbol ($QN^{(2)}$).}
\label{Q2}
\end{figure*}

In order to evaluate $\gamma_i$ and $\beta_{i,i'}$, we propose a new multi-class open queueing network ($QN^{(2)}$), comprised of $M/G/\infty$ nodes as in Fig. \ref{Q2}. In $QN^{(2)}$ each queueing node corresponds to a specific status of the buffer at the BS in the primary network at the beginning of a frame, similar to the states of the Markov chain presented in Section 3 (Fig. \ref{MC}). In other words, each queueing node indicates the corresponding number of empty slots in the current DL subframe. For example, node $u$ represents a DL subframe with $e^u=max⁡(M-u,0)$ total empty slots, where $u=0,1,...,N$.

The customers in this queueing network are the SU packets that should be transmitted in empty slots (after a contention is finished). In other words, arrivals at this queueing network are the packets arrived at node $TR$ of the first queueing network depicted in Fig. \ref{Q1}. So, the time symbols corresponding to the arrival instants of customers have been embedded in their classes.

The customers entering to $QN^{(2)}$ have class $(i,S_S)$, where $i$ denotes the time symbol when the packet transmission starts (it is important to note that here start of packet transmission means the end of contention among SUs, thus $i$ can be throughout the whole frame including the UL subframe, i.e., $1\leq i\leq K_{sym}$) and $S_S$ is the number of slots needed to transmit an SU packet. Therefore, in the queueing network in Fig. \ref{Q2} a customer of class $(x,s)$ arriving at node $u$ represents a packet whose transmission starts (or resumes when it is not transmitted completely in the previous DL subframe) at the $x$-th symbol in a frame with $e^u$ total empty slots in DL subframe while $s$ slots are needed to complete its transmission. If residual empty slots of the current frame for a customer with class $(x,s)$ $(e^u_{res}(x))$ are not sufficient to complete the packet transmission, the class of the customer becomes $(0,s-e_{res}^u(x))$ and is routed to another node. Class $(0,s-e_{res}^u(x))$ specifies that the corresponding packet transmission continues at the onset of the current frame. This customer is routed among the nodes in the queueing network until the required slots reaches zero.\\
\vspace{-0.08cm}
 Assume that $f^u(i)$, $i=1,2,...,\frac{K_{sym,DL}}{\nu}$, is a function that denotes the number of empty slots at the $i$-th column of DL subframe (see Fig. \ref{frame}). Obviously, for vertical striping, $f^u(.)$ for initial columns in which all slots are occupied equals zero. If $\nu$ is the number of time symbols in a slot and $K_{sym,DL}$ denotes the number of time symbols in a DL subframe, we define
\begin{align}\label{definel} 
& l^u_{(x,s)}=min_{\lceil \frac{x}{\nu}\rceil+1\leq j\leq \frac{K_{sym,DL}}{\nu}} \quad j,\\
& \nonumber \text{s.t.} \quad \sum_{i=\lceil \frac{x}{\nu}\rceil+1}^j f^u(i)-s\geq 0.
\end{align}
In fact $l^u_{(x,s)}$ indicates the number of the column of slots in which the transmission of the packet with class ($x,s$) in the current frame ends. Thus, the average service time of the customers with a typical class $(x,s)$ at node $u$ ($\tau^{u}_{(x,s)}$) in the case of horizontal and vertical stripings are obtained by
\begin{align}\label{horizontal} 
\begin{split}
\tau^{u}_{(x,s)}=
\left\{ 
\begin{array}{l l}
\scriptstyle T_{sym}\nu\left\lgroup l^u_{(x,s)}-(\lceil\frac{x}{\nu}\rceil +1)+\frac{s-\sum_{i=\lceil\frac{x}{\nu}\rceil+1}^{l^u_{(x,s)}-1}f^u(i)}{f^u(l^u_{(x,s)})}\right\rgroup;\\
\qquad \qquad \qquad \qquad \qquad    \scriptstyle \text{if $l^u_{(x,s)}$ exists}\\
\scriptstyle T_{frame}-xT_{sym}; \qquad \qquad \qquad  \text{o.w.}
\end{array}
\right.,
\end{split}
\end{align}
where $T_{sym}$ and $T_{frame}$ denote time duration of a time symbol and a WiMAX frame, respectively. Furthermore, in various rectangular striping algorithms in WiMAX, by assuming that empty slots of DL subframes are distributed uniformly over the entire DL subframe, the residual empty slots in the current frame ($e_{res}^u(x)$) and $\tau^{u}_{(x,s)}$ are given by ($1\leq x\leq K_{sym}$)
\begin{align}\label{rect} 
&e_{res}^u(x)=\frac{T_{DL}-xT_{sym}}{T_{DL}}e^u,\\
\label{TQ2-1}
&\tau^{u}_{(x,s)}=\left\{
\begin{array}{l l}
\frac{s}{e^u}T_{DL};&e_{res}^u(x)>s\\
T_{frame}-xT_{sym};&\text{o.w.}
\end{array}
\right.,
\end{align}
where $T_{DL}$ is time duration of a DL subframe. If $K_{sym,DL}=\frac{R}{R+1}K_{sym}$, where $R=\frac{T_{DL}}{T_{UL}}$ is the downlink to uplink subframe ratio, it is obvious that $K_{sym,DL}<x\leq K_{sym}$ denotes UL subframe and leads to $e_{res}^u(x)=0,\forall u$. Note that the UL interruptions through a transmission are considered in the service time of the queueing nodes. Routing probabilities of different customers turn out to be ($0\leq u,u'\leq N,0\leq x\leq K_{sym}$)
\begin{align}\label{RQ2-2}
&r^{u,out}_{(x,s),(x+\hat{\tau}_{(x,s)}^{u},0)}=\left\{
\begin{array}{l l}
1;&s<e_{res}^{u}(x)\\
0;&\text{o.w.}
\end{array}
\right. ,
\end{align}
\begin{align}\label{RQ2-3}
\begin{split}
&r^{u,u'}_{(x,s),(0,s-e_{res}^{u}(x))}=\left\{
\begin{array}{l l}
P_{u,u'};&e_{res}^{u}(x)<s\\
0;&\text{o.w.}
\end{array}
\right. ,\\
\end{split}
\end{align}
where $P_{u,u'}$ is the transition probability from state $u$ to state $u'$ in the Markov chain presented in Section 3, see (\ref{gbe2})-(\ref{gbe3}) and Fig. \ref{MC}. Therefore, the proposed model takes into account the fact that the number of empty slots in successive frames are dependent. As shown in Fig. \ref{Q2} customers of class $(i',0),i'=1,...,K_{sym,DL}$ (i.e., during a DL subframe) leave the network, because the transmission is completed at the $i'$-th OFDM symbol at the current DL subframe. \\
\vspace{-0.09cm}
The transmission time of a packet that its transmission has started at the $i$-th symbol, is the total residence time of the arriving customer with class $(i, S_S)$  in the queueing network (Fig. \ref{Q2}). Therefore, by using Little's law \cite{R21}, the average transmission time ($\gamma_i$ in (\ref{T15})) of such a typical packet is obtained by
\begin{equation}
\gamma_i=\sum_{u=0}^{N}\sum_{(x,s)} \frac{\alpha_{(x,s)}^u}{\lambda_{(i,S_S)}}\tau_{(x,s)}^u ,
\end{equation}
where $\alpha_{(x,s)}^u$ and $\lambda_{(i,S_S)}$ denote the arrival rate of customers of class $(x,s)$ at node $u$, and the arrival rate of customers of class $(i,S_S)$ at the queueing network, respectively. Meanwhile, note that ending a packet transmission and starting the next one (end of the next contention) are close (this time interval is shown with a dashed line in Fig. \ref{frame}) because control signals among SUs are transmitted in an independent control channel, so we assume that these two events occur in the same frame. Therefore, the routing probability of the arriving customers at the queueing network toward a typical queueing node (see Fig. \ref{Q2}), $r_{(i,S_S)}^{out,u}$, and the rate of customers departed from that node to the out of the queueing network, $\alpha_{(i',0)}^{u,out}$, which is equal to $\sum_{(i'',s'')}\alpha_{(i'',s'')}^u r_{(i'',s'' ),(i',0)}^{u,out}$, are related as
\begin{equation}
r^{out,u}_{(i,S_S)}=\sum_{i'=1}^{K_{sym,DL}}\frac{\alpha^{u,out}_{(i',0)}}{\lambda_{(i,S_S)}} ;\quad \scriptstyle 0\leq u\leq N.
\end{equation}
In order to find $\gamma_i$ and $\beta_{i,i'}$ for each packet transmission with different starting times, we have to solve traffic equations of $QN^{(2)}$ for each class $(i,S_S)$ of arriving customers separately. The traffic equations can be written as
\begin{equation}
\alpha^u_{(i,S_S)}=\lambda_{(i,S_S)}r^{out,u}_{(i,S_S)} ;\quad \scriptstyle 1\leq i\leq K_{sym},
\end{equation}
\begin{equation}
\alpha^u_{(0,s)}=\sum_{u'=0}^{N}\sum_{(x',s')}\alpha^{u'}_{(x',s')}r^{u',u}_{(x',s'),(0,s)}; \quad\scriptstyle 1\leq s\leq S_S,
\end{equation}
where $r^{u',u}_{(x',s'),(0,s)}$ is the probability that a customer of class $(x',s')$ is converted to class $(0,s)$ when departed from node $u'$ and routed to node $u$, i.e., the transmission has not ended at the current frame and it will resume in the next frame. Note that $\lambda_{(i,S_S)}$ is a dummy variable and its value does not affect the value of the desired parameters $\gamma_i$ and $\beta_{i,i'}$.

Moreover, the parameter $\beta_{i,i'}$ (the probability of ending a packet transmission at the $i'$-th symbol over the frame which has been started at the $i$-th symbol), is given by
\begin{equation}
\beta_{i,i'}=\frac{\sum_{u}\alpha^{u,out}_{(i',0)}}{\lambda_{(i,S_S)}} ;\quad \scriptstyle  1\leq i'\leq K_{sym}.
\end{equation}
Obviously, since no transmission ends in the UL subframe, $\beta_{i,i'}=0; \forall i'>K_{sym,DL}$. By solving the queueing network in Fig. \ref{Q1} and Fig. \ref{Q2}, we are able to solve (\ref{main}) and obtain the saturation throughput of the cognitive network.

It is worth noting that in order to include SU packet transmission in the UL opportunities in our proposed analytical approach as well, since the traffic characteristics of UL and DL subframes are different, we need to model the opportunities at UL and DL subframes, separately. In this respect, a two-dimensional Markov chain has to be used instead of the Markov chain in Section 3 to evaluate the empty slots in the DL and UL subframes. However, the overall structure of $QN^{(1)}$ which represents the packet transmission process of a secondary user remains the same and only the parameters related to packet transmission time (service times and routing probabilities) would change. Moreover, some new nodes are added to $QN^{(2)}$ which are used to estimate the SUs’ packet transmission parameters in DL and UL subframes. The new nodes correspond to the second dimension of the proposed Markov chain for the primary network. As indicated before, for the sake of simplicity and importance of DL subframes, we ignore exploiting opportunities in UL subframes.

It is important to note that a similar approach with suitable parameters could be used to analytically evaluate the saturation throughput of the secondary users in the presence of an LTE primary network as well. Considering that OFDMA is used in LTE downlink subframes too \cite{R5}, the general approach of estimating the saturation throughput in this paper would not change. The LTE BS takes scheduling decisions at starting of each subframe and PDCCH (physical downlink control channel) carries scheduling assignment information for users (like DL/UL MAP in WiMAX). With a specific downlink/uplink configuration in LTE \cite{R4}, the parameter values are considered accordingly.

\section{Numerical Results}
In this section we employ our analytical model and derive the saturation throughput in different conditions. To support our analysis, simulation results are also reported. The parameters used for primary and secondary networks are listed in Table \ref{parameter}. 

\begin{table}[t]
  \small
  \centering
  \caption{Typical Parameter Values}
    \begin{tabular}{"c|c"}
    \thickhline
    \textbf{\quad Parameter} & \textbf{Value} \\
    \thickhline
    \quad  $T_{frame}$ & 5 ($ms$)  \\\hline
    \quad Number of subchannels & 30 \\\hline 
		\quad $K_{sym,DL}$ & 30,26,14 \\\hline 
		\quad Frequency bandwidth & 10 ($MHz$) \\\hline 
		\quad $R$ & $ \frac{13}{12}, \frac{3}{2}, \frac{4}{1}$ \\\hline
		\quad $m$ & 4 \\\hline 
		\quad $S_P$ & 10 \\\hline 
		\quad $S_S$ & 60 \\\hline 
		\quad $C_B$ & 55 ($packets$) \\\hline
   \quad $W_0$ & 4 \\\hline 
    \end{tabular}%
  \label{parameter}%
\end{table}%

The simulation of the network scenario has been done in MATLAB environment. There are $N_s$ WLAN nodes as secondary users. Each node always has packets in its buffer to send to AP (i.e., it is saturated). They use IEEE 802.11 MAC protocol to contend along the time. The winner node starts packet transmission in the empty slots of the WiMAX DL subframe. Simulation of the primary network is similar to the analytical model described in the paper. So, there is a finite buffer where packets arrive with exponential inter-arrival time and they are blocked if the buffer is full. Packets in the buffer are scheduled to be sent in the coming DL subframe with a finite number of slots. Horizontal striping algorithm is used as the usual resource unit allocation algorithm in simulations.

\begin{figure}
\centering
\includegraphics[width=2.7in]{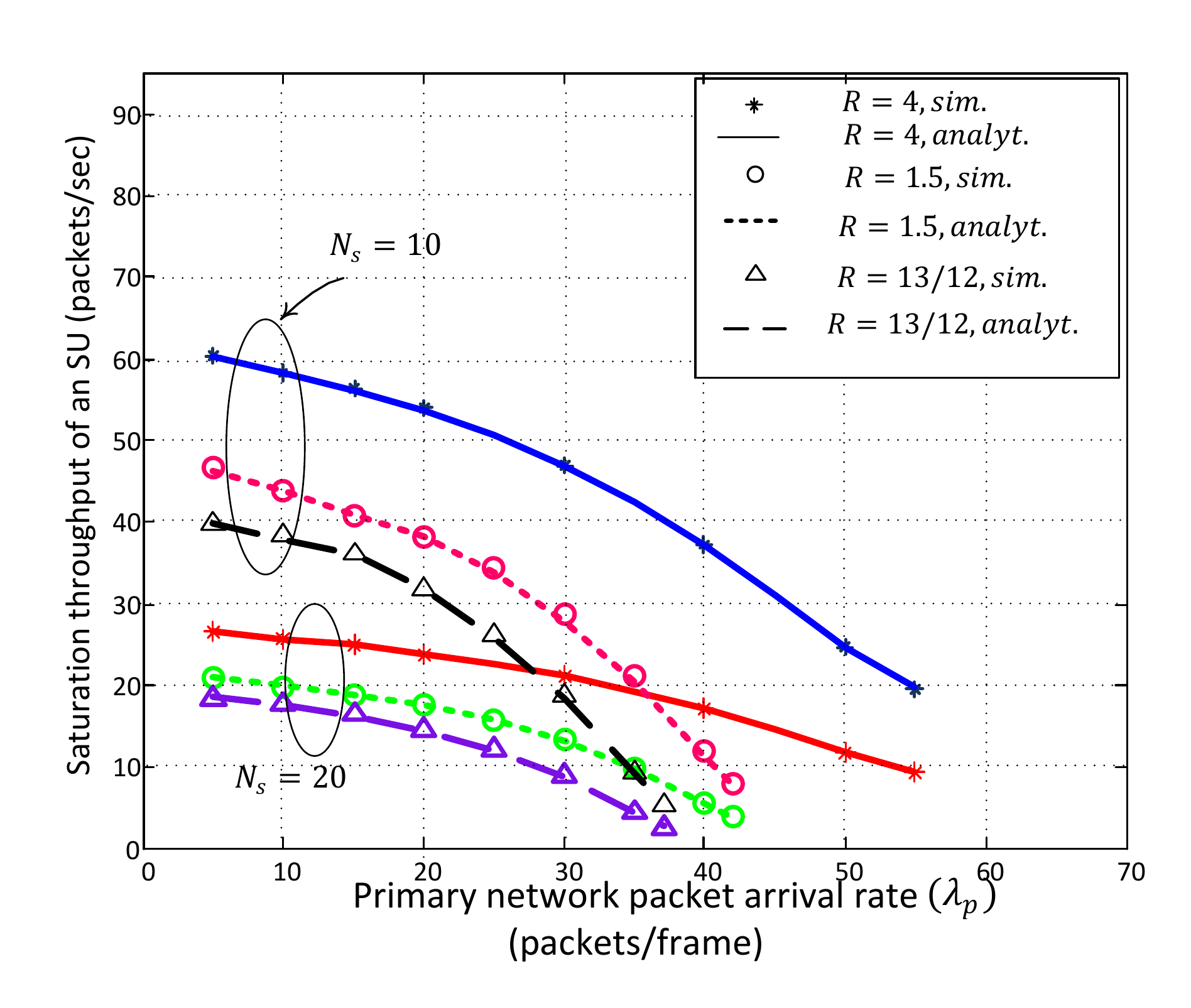}
\caption{Saturation throughput of an SU versus primary network packet arrival rate.}
\vspace{-0.6cm}
\label{fig1r}
\end{figure}

The reduction in saturation throughput of an SU with increase of the packet arrival rate in the primary network is observed in Fig. \ref{fig1r}. In this plot, three different downlink to uplink subframe ratios ($R$) are considered ($R=\frac{3}{2},\frac{13}{12},\frac{4}{1}$). In this figure, two sets of the results corresponding to two number of secondary nodes ($N_s=10,20$) have been shown. As we observe in the figure, when the DL to UL ratio is larger (e.g., 4), more resources have been dedicated to downlink, so the decrease rate of saturation throughput of SUs is smaller compared with lower DL to UL ratio (e.g., 1.5). 

\begin{figure}
\centering
\includegraphics[width=2.7in]{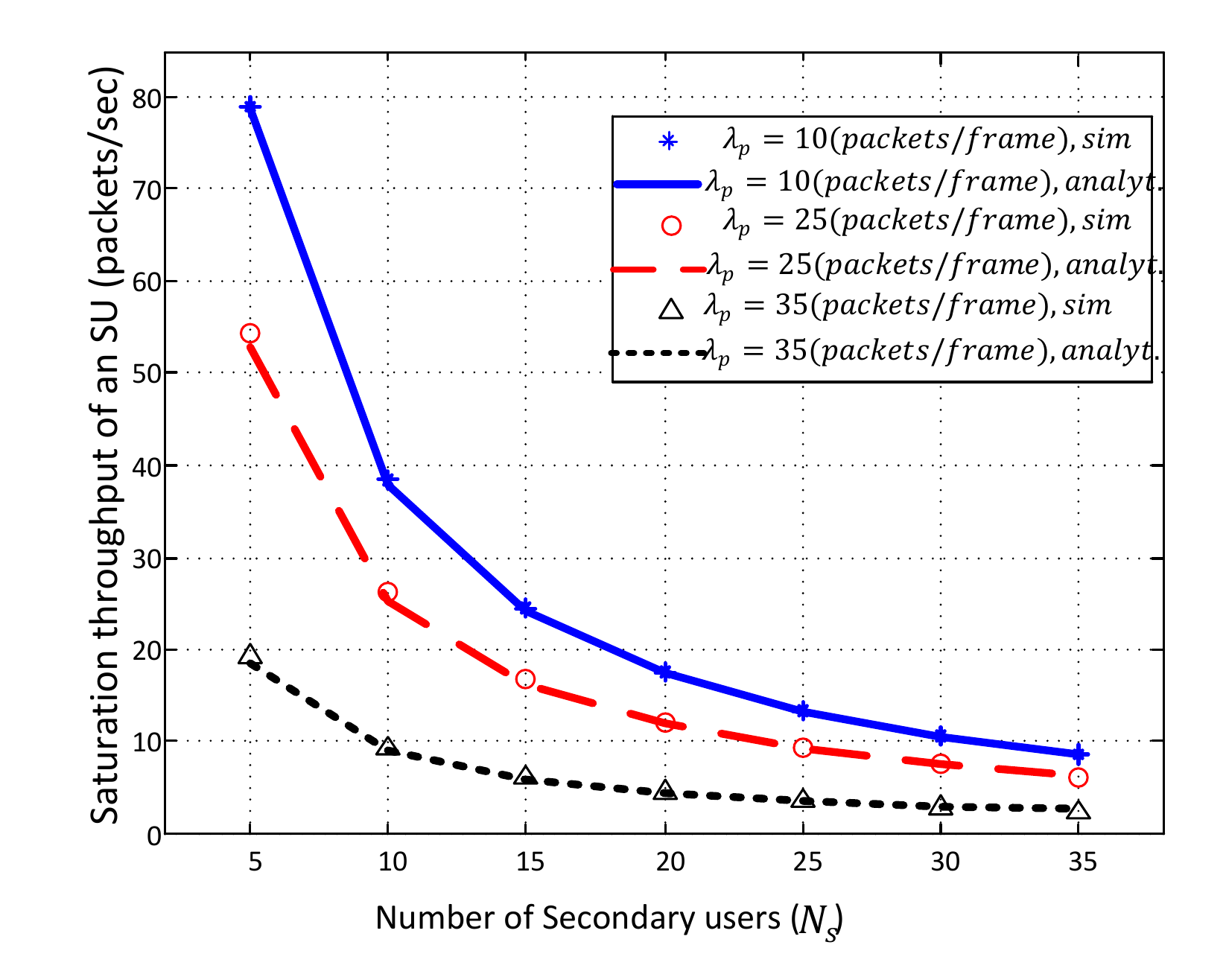}
\caption{Saturation throughput of an SU versus the number of secondary nodes ($R=\frac{13}{12}$).}
\label{fig2r}
\vspace{-0.4cm}
\end{figure}

\begin{figure}
\centering
\includegraphics[width=2.7in]{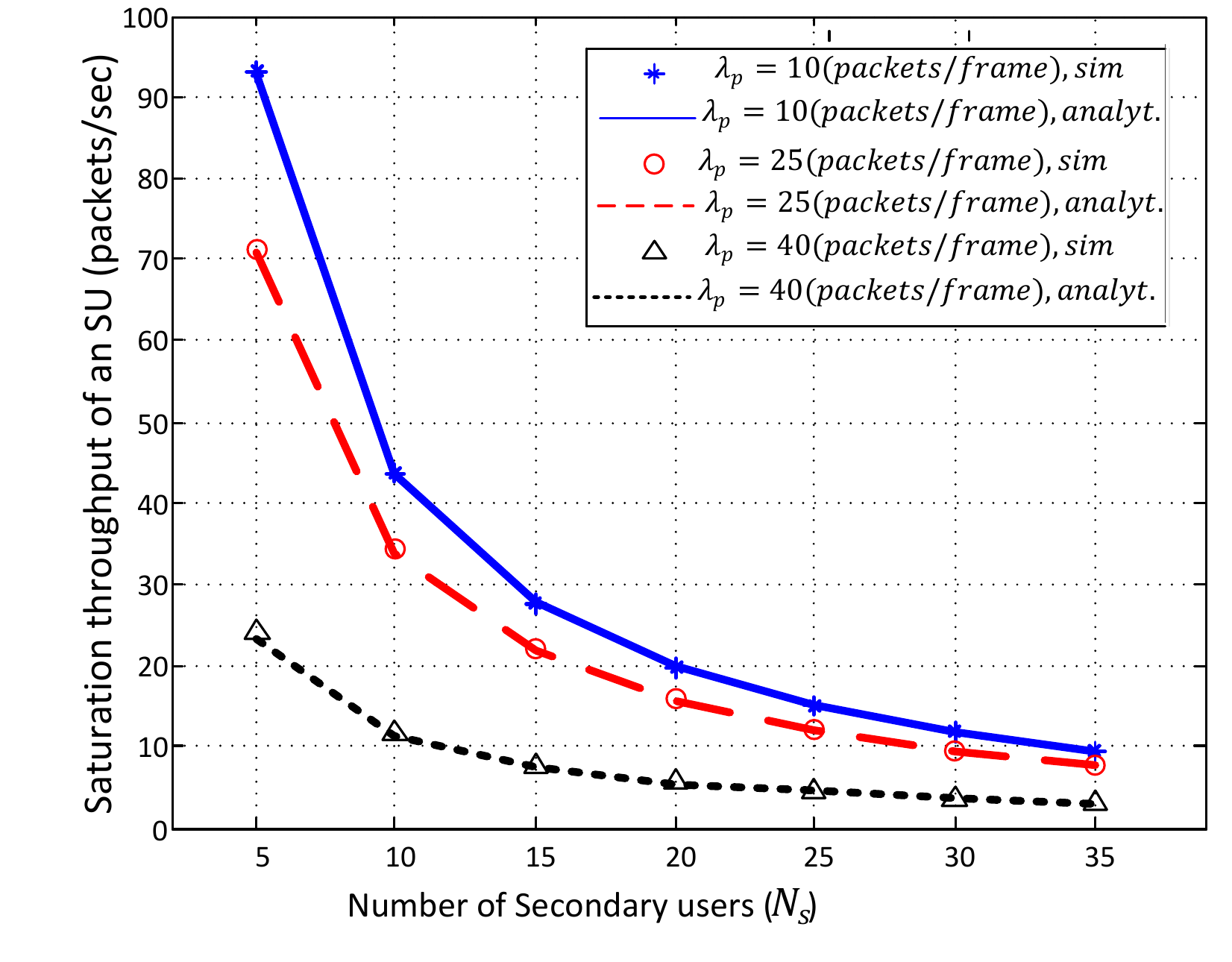}
\caption{Saturation throughput of an SU versus the number of secondary nodes ($R=\frac{3}{2}$).}
\label{fig3r}
\end{figure}

\begin{figure}
\centering
\includegraphics[width=2.7in]{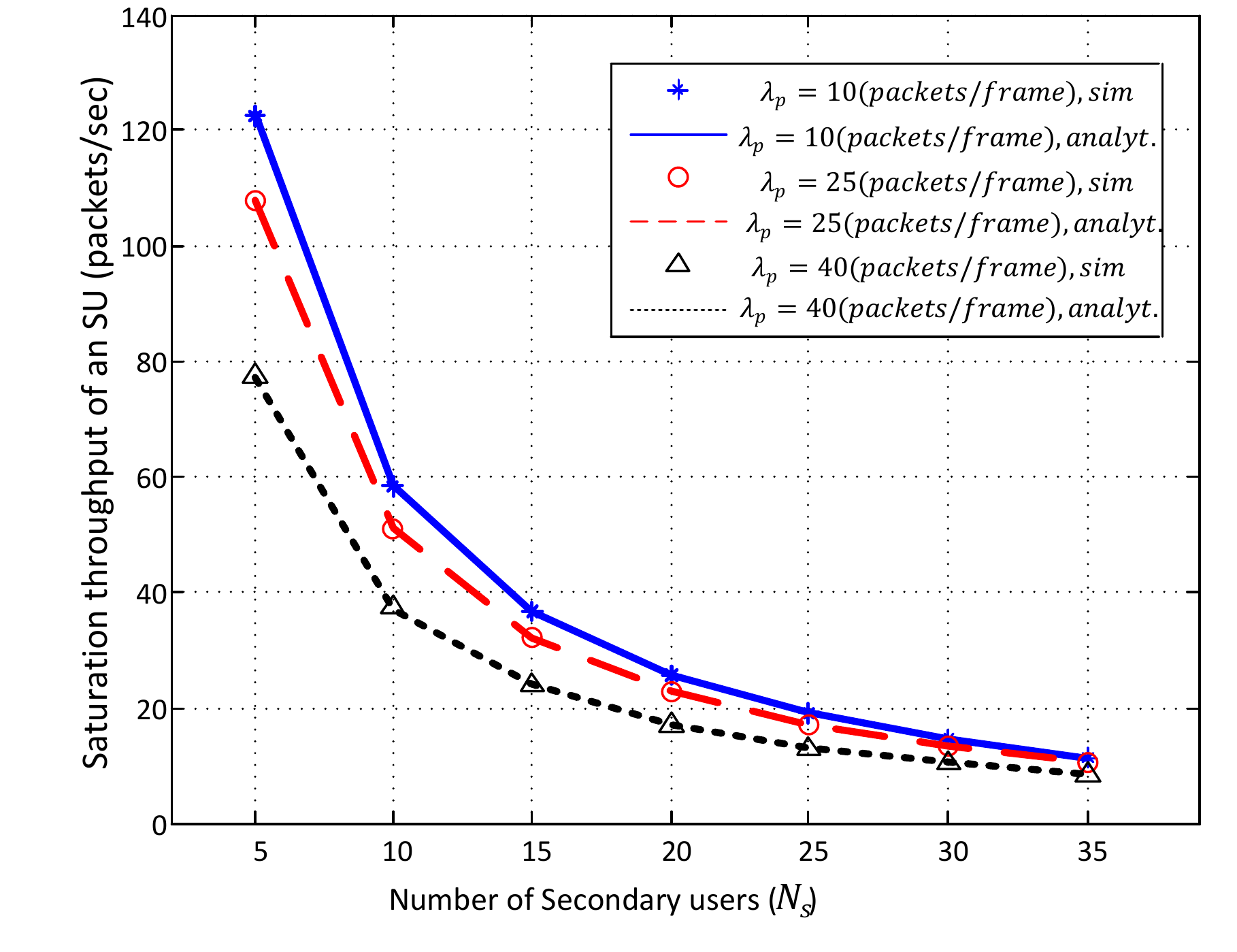}
\caption{Saturation throughput of an SU versus the number of secondary nodes ($R=\frac{4}{1}$).}
\label{fig4r}
\end{figure}

In Figs. \ref{fig2r}, \ref{fig3r}, and \ref{fig4r} the saturation throughput of an SU is plotted versus the number of SUs with $R=\frac{13}{12}$, $R=\frac{3}{2}$, and $R=\frac{4}{1}$, respectively. The maximum number of SUs with a specified minimum desired saturation throughput is obtained from these figures. For example to have a minimum saturation throughput of 20 (\emph{packets/sec}) in each SU, the number of users in the secondary network should be less than 5, 13, and 18, with primary network arrival rate 35, 25, and 10 (\emph{packets/frame}), respectively ($R=\frac{13}{12}$). However, for the last case (i.e., $\lambda_p=10$ (\emph{packets/frame})) the corresponding number of users in the secondary network should be less than 20 and 25 with $R=\frac{3}{2}$ and $R=\frac{4}{1}$, respectively. By observing Figs. \ref{fig2r}, \ref{fig3r}, and \ref{fig4r}, we infer that the number of maximum number of SUs supported with a specific minimum saturation throughput has a saturating behavior versus $R$, i.e., by increasing $R$, we do not observe a significant increase in the number of SUs. 

\begin{figure}
\centering
\includegraphics[width=2.7in]{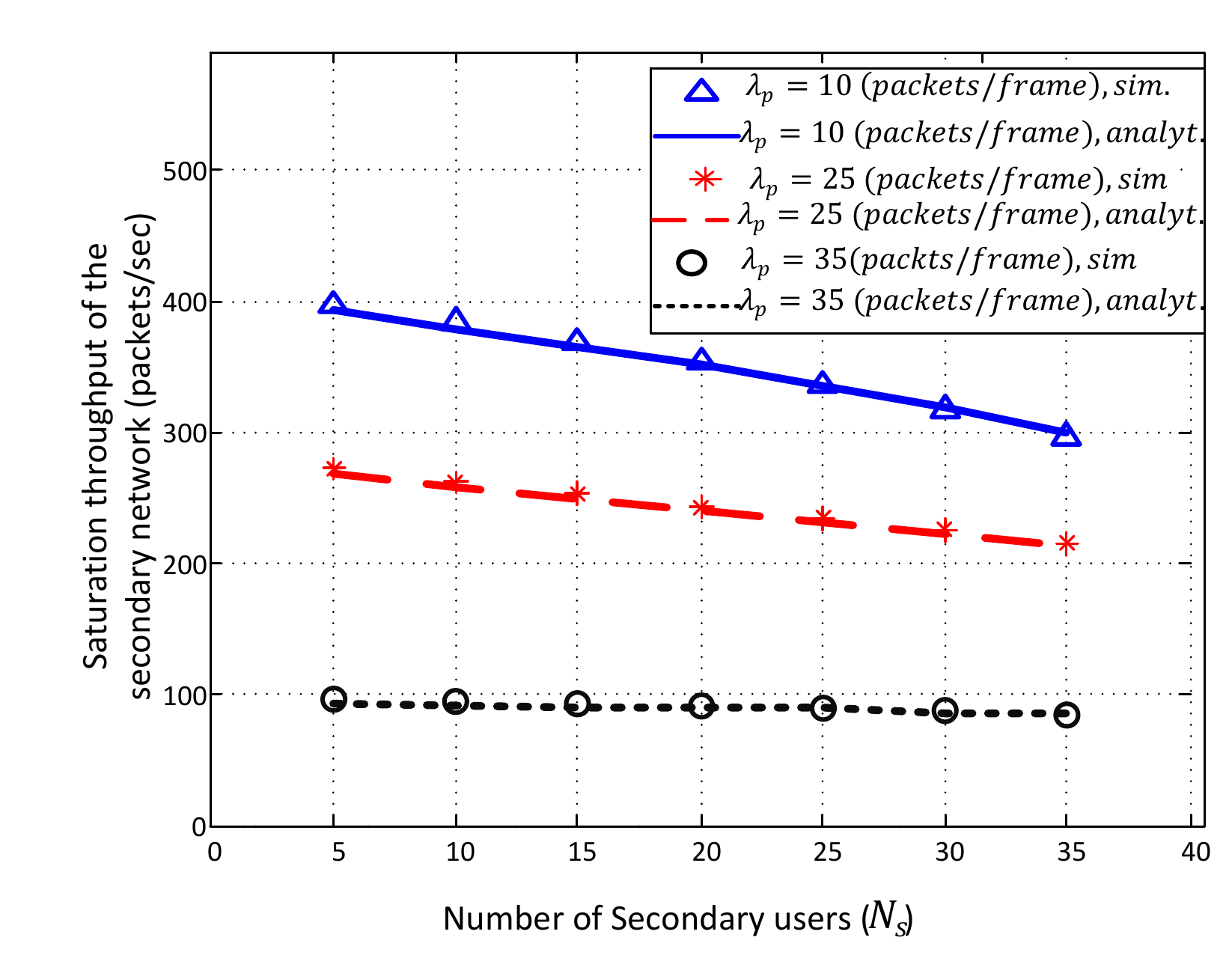}
\caption{Saturation throughput of the secondary network versus the number of secondary nodes ($R=\frac{13}{12}$).}
\vspace{-0.5cm}
\label{fig5r}
\end{figure}

Fig. \ref{fig5r} depicts the saturation throughput of the secondary network versus the number of secondary nodes. It can be seen that as the number of SUs increases, the total saturation throughput of the secondary network decreases due to more collisions among SUs. However, when the WiMAX frames are heavily loaded, we observe that the total saturation throughput of the secondary network remains almost the same irrespective of the number of SUs. Since in this situation, the length of colliding interval compared to the length of successful packet transmission becomes smaller,the throughput increase due to larger number of SUs compensates the degrading effect of the increased probability of collision.


\begin{figure}
\centering
\includegraphics[width=2.7in]{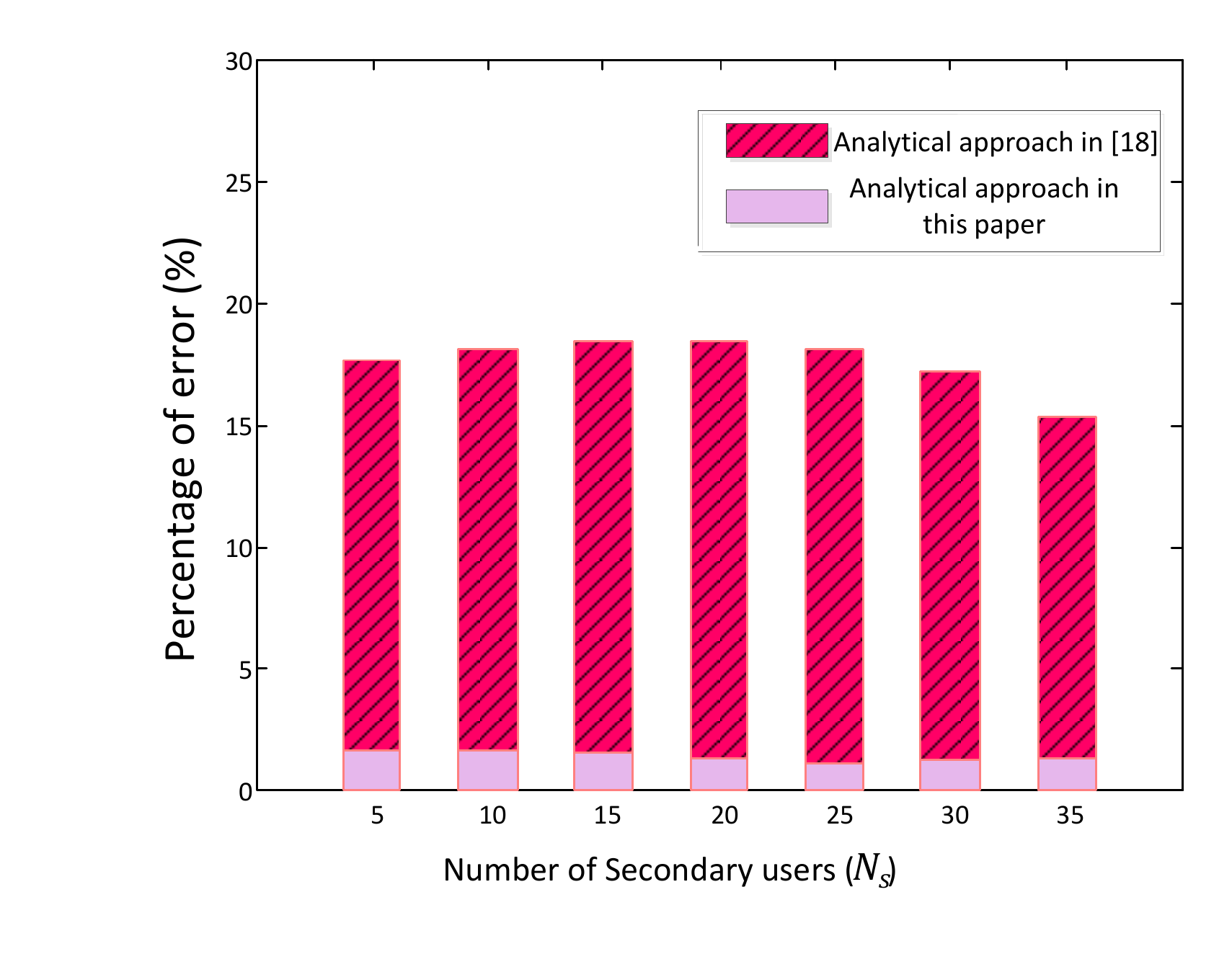}
\caption{Percentage of error in two analytical models ($\lambda_p=25$ (packets/frame), $R=\frac{13}{12}$).}
\vspace{-0.5cm}
\label{fig6r}
\end{figure}

To show the increased accuracy in our approach compared with simplified analytical approach in \cite{R28}, the amount of mismatch between analytical and simulation results for saturation throughput of an SU has been illustrated in Fig. \ref{fig6r}. As we observe by the proposed analytical approach in this paper, we have reduced the mismatch error from more than 18\% to less than 3\%. Thus, the detailed analysis of the scenario is very effective in more accurate evaluation of the saturation throughput. Roughly speaking, this high mismatch error in \cite{R28} could lead to about 20\% error in determining the maximum number of SUs to have a minimum required saturation throughput.
\begin{figure}
\centering
\includegraphics[width=2.7in]{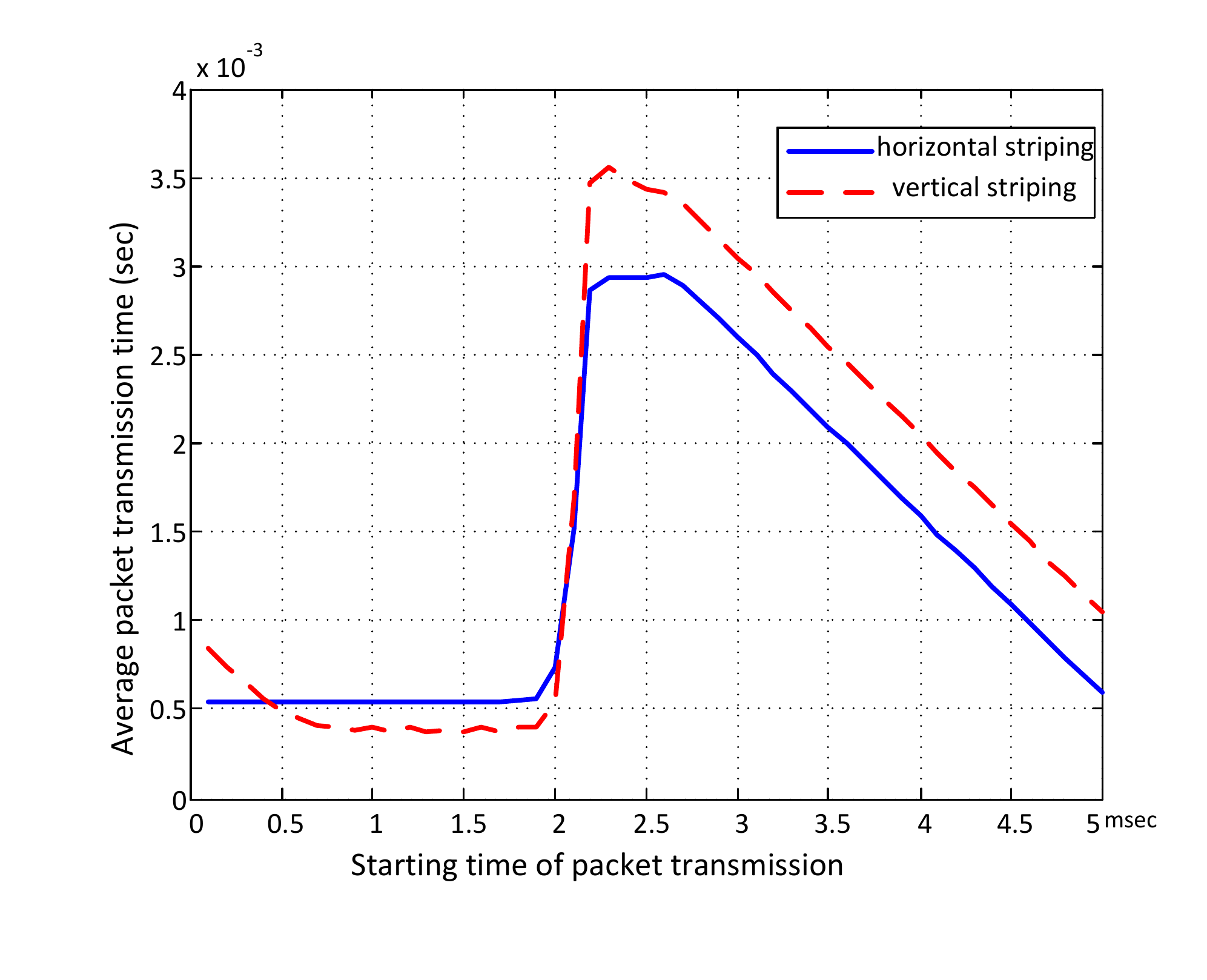}
\caption{The average packet transmission times of an SU with different starting times along the frame for horizontal and vertical striping ($R=\frac{13}{12}$). }
\label{fig7r}
\end{figure}  
\begin{figure}
\vspace{-0.3cm}
\centering
\includegraphics[width=2.7in]{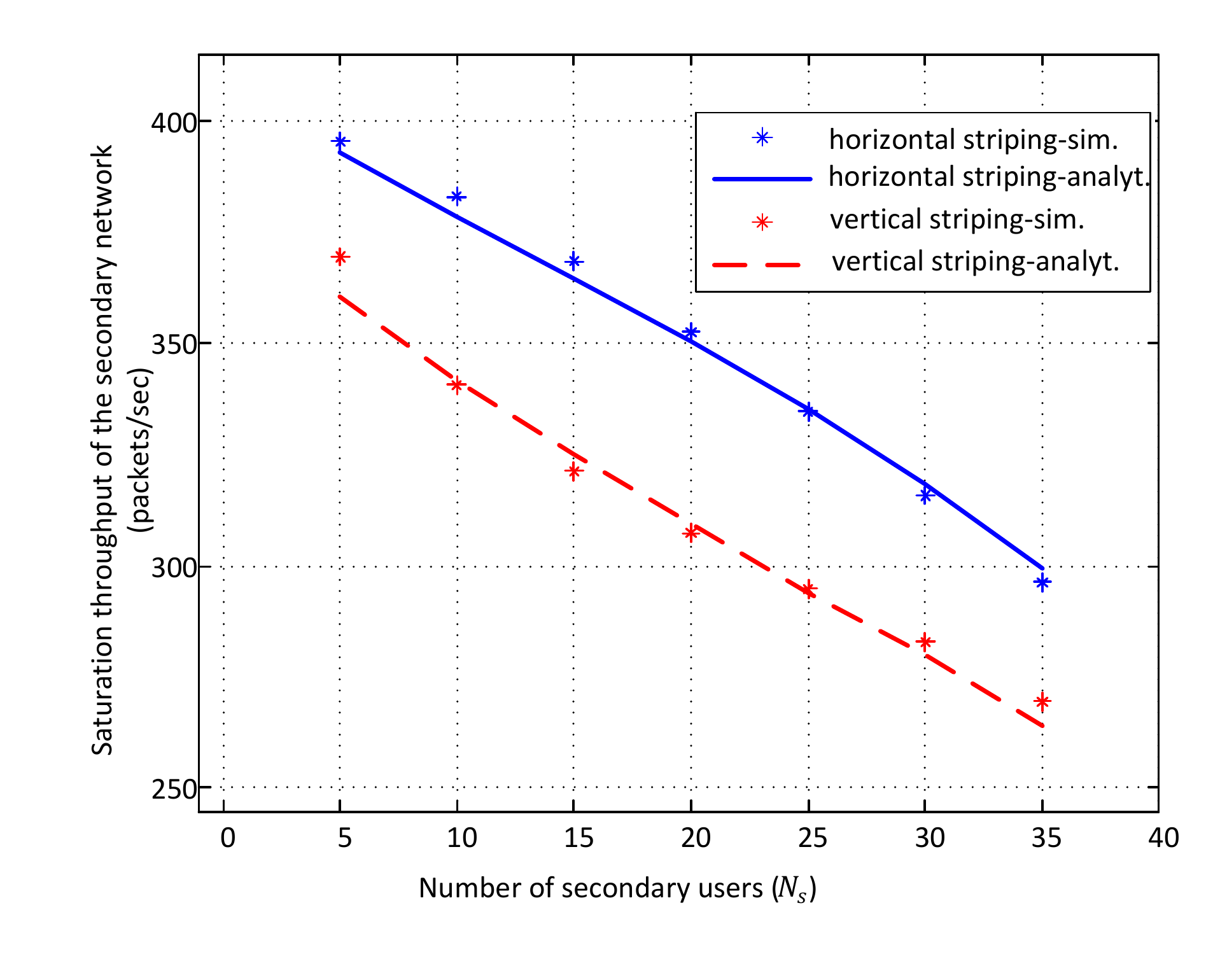}
\caption{Saturation throughput of an SU versus the number of secondary nodes for horizontal and vertical striping algorithms ($R=\frac{13}{12}$).}
\vspace{-0.5cm}
\label{fig8r}
\end{figure}
\begin{figure}
\centering
\includegraphics[width=2.7in]{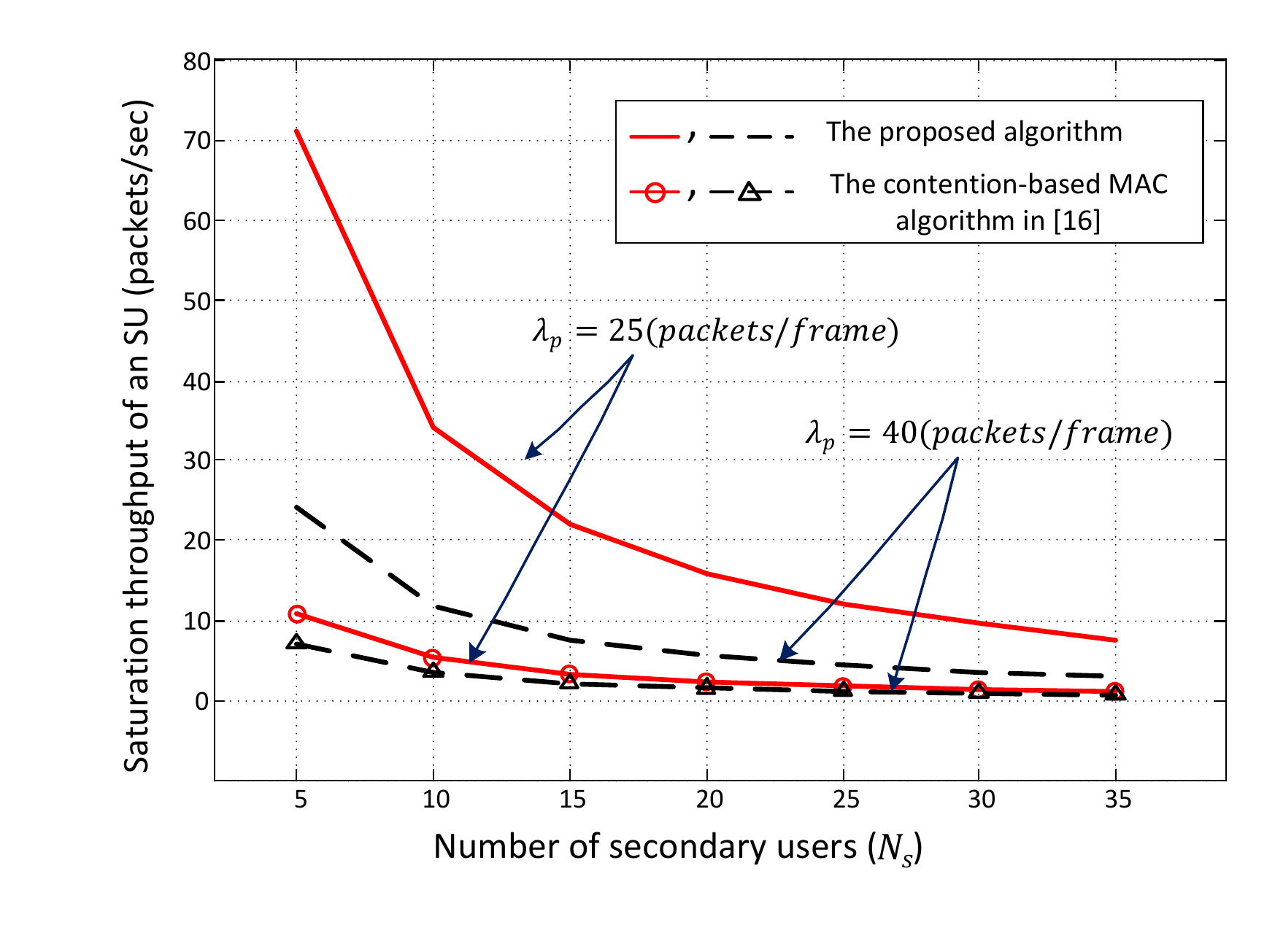}
\caption{Saturation throughput of an SU of the proposed algorithm and the contention-based MAC algorithm proposed in \cite{R26} ($R=\frac{3}{2}$).}
\label{fig9r}
\end{figure}

In Fig. \ref{fig7r} the average packet transmission times of an SU with different starting times are plotted with $R=\frac{13}{12}$ and $\lambda_p=25 (packets/frame)$ for horizontal and vertical striping algorithms for WiMAX. Since it is more likely that there is no empty slots in the initial columns of DL subframe in the vertical striping, it is observed that average transmission time for packet transmissions started in the beginning of DL subframes (up to 0.5 \emph{msec} approximately, in Fig. \ref{fig7r}), is more compared to the case of horizontal striping. While for the next few columns (0.5-2 \emph{msec} in Fig. \ref{fig7r}), since all the slots are empty in the vertical striping algorithm, the average packet transmission time is less in comparison with the horizontal striping case. Furthermore, in columns near or during the UL subframe (2-5 \emph{msec} in Fig. \ref{fig7r}), since the initial columns of slots of the next DL subframe in the vertical striping algorithm are fully occupied, packet transmissions last further in average compared to horizontal striping. It seems that if the opportunities of primary network are sporadic, the packet transmission time for SUs is shorter in average. Fig. \ref{fig8r} confirms this conclusion, as the saturation throughput for horizontal striping case is more than saturation throughput when vertical striping algorithm is used.

Moreover, we have compared the saturation throughput of an SU in our considered scenario with an algorithm similar to the contention-based MAC algorithm proposed in \cite{R26}, where SUs wait a random number of exploitable WiMAX frames with exponential backoff procedure. The results are shown in Fig. \ref{fig9r} for two values of $\lambda_p$. As explained before, because of independency of the time slots of our 802.11-based MAC algorithm with the frame structure of WiMAX, exploiting the empty slots of WiMAX is more efficient in our scenario, with a cost of having a separate control channel.

\section{Conclusion}
Since the primary users and the resource allocation in the primary network render opportunities for secondary users, exact modeling of these issues is very important in cognitive network evaluation. We considered a WLAN based on IEEE 802.11 MAC protocol overlaid on a primary WiMAX in TDD mode. By using a Markov chain the number of empty slots at different frames of WiMAX were probabilistically obtained. Then we employed an open queueing network for evaluating the packet transmission process of the secondary network. In order to derive the saturation throughput as one of the important performance metrics, we needed to compute the average service time and routing probabilities at different queueing nodes. Regarding random number of empty slots at different frames and dependency between the number of empty slots at consecutive frames, we proposed a multi-class open queueing network that enabled us to derive the average transmission time of a typical secondary packet as well as the probability that a packet transmission will finish at each moment along a frame. Then we derived the saturation throughput of the secondary network in different conditions. We also confirmed our analytical results by simulation. The numerical results showed high accuracy of the proposed analytical approach. Furthermore the comparison between the results for vertical and horizontal striping algorithms showed that if opportunities in the primary network are sporadic, the saturation throughput of SUs is improved.

\bibliographystyle{IEEEtran}
\bibliography{references}

\ifCLASSOPTIONcaptionsoff
  \newpage
\fi

\end{document}